\documentclass[12pt]{article}
\usepackage{amsmath}
\usepackage{amsfonts}
\usepackage{cprotect}
\usepackage{graphicx}
\usepackage{enumerate}
\usepackage{natbib}
\usepackage{booktabs}
\usepackage[flushleft]{threeparttable}
\usepackage{url} 

\newcommand{\blind}{1}

\begin{document}

\bibliographystyle{apalike}

\def\spacingset#1{\renewcommand{\baselinestretch}%
{#1}\small\normalsize} \spacingset{1}


\if1\blind
{
  \title{\bf Longitudinal Modeling of Age-Dependent Latent Traits with Generalized Additive Latent and Mixed Models}
  \author{{\O}ystein S{\o}rensen\thanks{
    The authors gratefully acknowledge \textit{the European Research Council under grant agreements 283634, 725025 (to A.M.F.) and 313440 (to K.B.W.), the Norwegian Research Council (to A.M.F., K.B.W.), The National Association for Public Health’s dementia research program, Norway (to A.M.F) and center support from the University of Oslo}}\hspace{.2cm}\\
    Department of Psychology, University of Oslo\\
    and \\
    Anders M. Fjell \\
    Department of Psychology, University of Oslo\\
    Department of Radiology and Nuclear Medicine, Oslo University Hospital\\
    and \\
    Kristine B. Walhovd \\
    Department of Psychology, University of Oslo\\
    Department of Radiology and Nuclear Medicine, Oslo University Hospital}
  \maketitle
} \fi

\if0\blind
{
  \bigskip
  \bigskip
  \bigskip
  \begin{center}
    {\LARGE\bf Longitudinal Modeling of Age-Dependent Latent Traits with Generalized Additive Latent and Mixed Models}
\end{center}
  \medskip
} \fi

\bigskip
\begin{abstract}
We present generalized additive latent and mixed models (GALAMMs) for analysis of clustered data with responses and latent variables depending smoothly on observed variables. A scalable maximum likelihood estimation algorithm is proposed, utilizing the Laplace approximation, sparse matrix computation, and automatic differentiation. Mixed response types, heteroscedasticity, and crossed random effects are naturally incorporated into the framework. The models developed were motivated by applications in cognitive neuroscience, and two case studies are presented. First, we show how GALAMMs can jointly model the complex lifespan trajectories of episodic memory, working memory, and speed/executive function, measured by the California Verbal Learning Test (CVLT), digit span tests, and Stroop tests, respectively. Next, we study the effect of socioeconomic status on brain structure, using data on education and income together with hippocampal volumes estimated by magnetic resonance imaging. By combining semiparametric estimation with latent variable modeling, GALAMMs allow a more realistic representation of how brain and cognition vary across the lifespan, while simultaneously estimating latent traits from measured items. Simulation experiments suggest that model estimates are accurate even with moderate sample sizes.
\end{abstract}

\noindent%
{\it Keywords:}  brain and cognition, generalized additive mixed models, latent variable modeling, lifespan trajectories, mixed response.
\vfill

\newpage
\spacingset{1} 
\section{Introduction}
\label{sec:introduction}

Generalized linear mixed models (GLMMs) and nonlinear mixed models are widely used whenever observations can be divided into meaningful clusters. However, they require the parametric form of the effects to be exactly specified, and in many applications this may be impractical or not possible. For example, when studying how the human brain changes over the lifespan, volumes of different brain regions exhibit distinctive trajectories, differing with respect to rate of increase during childhood, age at which maximum is attained, and rate of decline in old age \citep{bethlehemBrainChartsHuman2022,sorensenRecipeAccurateEstimation2021}. Similarly, domain-specific cognitive abilities follow unique lifespan trajectories, with traits like episodic memory and processing speed peaking in early adulthood, while acquired knowledge like vocabulary peaks in late adulthood \citep{mcardleComparativeLongitudinalStructural2002,tucker-drobCognitiveAgingDementia2019}. Generalized additive mixed models (GAMMs) \citep{woodGeneralizedAdditiveModels2017a} flexibly estimate nonlinear relationships by a linear combination of known basis functions subject to smoothing penalty, and are ideally suited to these applications. 

Both GLMMs and GAMMs can be used for analyzing multivariate response data, allowing estimation of correlated change across multiple processes. However, when multivariate responses are considered noisy realizations of lower-dimensional latent variables, GLMMs and GAMMs essentially assume a parallel measurement model \citep{novickAxiomsPrincipalResults1966}, in which the coefficients relating latent to observed variables are known at fixed values. Structural equation models (SEMs) offer more flexible latent variable modeling, and extensions of the SEM framework include nonlinear models \citep{armingerBayesianApproachNonlinear1998,leeStatisticalAnalysisNonlinear2000}, latent basis models \citep{meredithLatentCurveAnalysis1990}, random forests \citep{brandmaierTheoryguidedExplorationStructural2016,brandmaierRecursivePartitioningContinuous2018} and models for categorical and ordinal response data \citep{muthenGeneralStructuralEquation1984}. Despite these advances, use of SEMs can be impractical when analyzing multilevel unstructured data, with explanatory variables varying at different levels \citep{curranHaveMultilevelModels2003}. Several proposed models bring SEMs closer to the flexibility of GLMMs, while retaining their ability to model latent variables \citep{driverContinuousTimeStructural2017,driverHierarchicalBayesianContinuous2018,mehtaPeopleAreVariables2005,mehtaPuttingIndividualBack2000,muthenSEMGeneralLatent2002,oudContinuousTimeState2000,proust-limaAnalysisMultivariateMixed2013,proust-limaEstimationExtendedMixed2017,rabe-heskethGeneralizedMultilevelStructural2004}. In particular, generalized linear latent and mixed models (GLLAMMs) \citep{rabe-heskethGeneralizedMultilevelStructural2004,skrondalGeneralizedLatentVariable2004} exploit the equivalence between random effects and latent variables \citep{skrondalLatentVariableModelling2007} to model latent and explanatory variables varying at any level. GLLAMMs are nonlinear hierarchical models whose marginal likelihood can be approximated by numerical integration over the latent variables \citep{rabe-heskethMaximumLikelihoodEstimation2005}. As GLLAMMs model the observed responses with an exponential family distribution, they are not limited to factor analytic measurement models, and incorporate important psychometric methods like item response models and latent class models.

While nonlinear modeling is possible with GLLAMMs, as with GLMMs the functional parametric forms are assumed known. In this paper, we introduce generalized additive latent and mixed models (GALAMMs), a semiparametric extension of GLLAMMs in which both the linear predictor and latent variables may depend smoothly on observed variables. Utilizing the mixed model view of smoothing \citep{kimeldorfCorrespondenceBayesianEstimation1970,ruppertSemiparametricRegression2003,silvermanAspectsSplineSmoothing1985,woodGeneralizedAdditiveModels2017a}, we show that any GALAMM can be represented as a GLLAMM, with smoothing parameters estimated by maximum marginal likelihood. Next, we show how a Laplace approximation to marginal likelihood of GLLAMMs can be computed efficiently using direct sparse matrix methods \citep{davisDirectMethodsSparse2006}, and maximized using the limited-memory Broyden-Fletcher-Goldfarb-Shanno algorithm with box constraints (L-BFGS-B) \citep{byrdLimitedMemoryAlgorithm1995} with gradients computed by automatic differentiation \citep{baydinAutomaticDifferentiationMachine2018}.

The proposed methods are similar to fully Bayesian approaches to semiparametric latent variable modeling \citep{fahrmeirBayesianSemiparametricLatent2007,songSemiparametricLatentVariable2010,songBayesianModelingApproach2013,songBayesianSemiparametricDynamic2013,songLatentVariableModels2014}, all of which have been limited to latent variables varying at a single level. In contrast, GALAMMs allow any number of levels, and due to the use of sparse matrix methods, crossed random effects are easily accommodated. A related Bayesian approach to semiparametric latent variable modeling has been based on finite mixture models \citep{bauerSemiparametricApproachModeling2005,kelavaGeneralNonlinearMultilevel2014,kelavaNonlinearStructuralEquation2014,yangBayesianSemiparametricStructural2010}.

The paper proceeds as follows. In Section \ref{sec:Background} we give brief introductions to GAMMs and GLLAMMs. In Section \ref{sec:models} we start by presenting the proposed framework, then show how GALAMMs can be represented as GLLAMMs with an additional level of latent variables corresponding to penalized spline coefficients. In Section \ref{sec:Estimation} we propose an algorithm for maximum marginal likelihood estimation of the models. In Section \ref{sec:cog_app} we present an example application illustrating how lifespan trajectories of abilities in three cognitive domains can be estimated from longitudinal data, combining the results of multiple tests taken at each timepoint. Next, in Section \ref{sec:ses_model} we study how socioeconomic status is associated with hippocampal volume across the lifespan. Each application example is followed by simulation experiments in Sections \ref{sec:cog_sim} and \ref{sec:ses_simulation}, respectively, closely following the data structure and parameters of the real data analysis. We discuss the results in Section \ref{sec:discussion}, and conclude in Section \ref{sec:conclusion}.

\section{Background}
\label{sec:Background}

Before presenting the proposed model framework, we start by providing brief background on its two major components, generalized additive models (GAMs) \citep{hastieGeneralizedAdditiveModels1986} and GLLAMMs. Along the way we also introduce the notation used in the paper.

\subsection{Generalized additive models as mixed models}
\label{sec:GAM_intro}

We here show how GAMs can be represented as mixed models, considering a model with a single univariate term for ease of exposition. The extension to multiple smooth terms or multivariate terms \citep{woodLowRankScaleInvariantTensor2006,woodStraightforwardIntermediateRank2013} follows the same steps. The ideas date back to \citet{kimeldorfCorrespondenceBayesianEstimation1970}, and have been presented in various forms since then \citep{linInferenceGeneralizedAdditive1999,silvermanAspectsSplineSmoothing1985,speedThatBLUPGood1991,verbylaAnalysisDesignedExperiments1999,woodStableEfficientMultiple2004,woodFastStableRestricted2011}. For an introduction to GAMs, we refer to the books \citet{ruppertSemiparametricRegression2003} and \citet{woodGeneralizedAdditiveModels2017a}.

Consider $n$ responses $y_{1}, \dots, y_{n}$, independently distributed according to an exponential family with density
\begin{equation}
\label{eq:galamm_distribution}
    f\left(y | \theta, \phi\right) = \exp \left( \frac{y\theta - b\left(\theta(\mu)\right)}{\phi} + c\left(y, \phi\right) \right)
\end{equation}
where $\mu = g^{-1}(\nu)$ is the mean, $g^{-1}(\cdot)$ is  the inverse of link function $g(\cdot)$, $\nu$ is a linear predictor, $\phi$ is a dispersion parameter, and $b(\cdot)$ and $c(\cdot)$ are known functions. For ease of exposition we consider a canonical link function, so $\theta(\cdot) = g(\cdot)$ and thus $\theta(\mu) = \theta(g^{-1}(\nu)) = \nu$. GAMs model the effect of a variable $x$ on the linear predictor with a function $f(x)$, constructed as a weighted sum of $K$ basis functions, $b_{1}(x), \dots, b_{K}(x)$ with weights $\beta_{1},\dots,\beta_{K}$. In the intermediate rank approach to smoothing \citep{woodFastStableRestricted2011} the basis functions are regression splines, and the number of basis functions is much smaller than the sample size, while still being large enough to represent a wide range of function shapes. Possible basis functions for the methods discussed in this paper include cubic regression splines \citep[Ch. 5.3.1]{woodGeneralizedAdditiveModels2017a}, P-splines \citep{eilersFlexibleSmoothingBsplines1996}, thin-plate regression splines \citep{woodThinPlateRegression2003}, and quadratic spline bases \citep[Ch. 3.6]{ruppertSemiparametricRegression2003}. 

In matrix-vector notation, with $\mathbf{y} = [y_{1}, \dots, y_{n}]^{T}$, $\boldsymbol{\beta} = [\beta_{1}, \dots, \beta_{K}]^{T}$, and $\mathbf{X} \in \mathbb{R}^{n \times K}$ with elements $X_{ij} = b_{j}(x_{i})$ for $i=1,\dots,n$ and $j=1,\dots,K$, the linear predictor is $\boldsymbol{\nu} = \mathbf{X}\boldsymbol{\beta}$, which together with \eqref{eq:galamm_distribution} defines a generalized linear model (GLM). We also assume that $f(x)$ is smooth, as measured by the integral of its squared second derivative over $\mathbb{R}$, which can be written $\int f''(x)^{2} \text{d} x = \boldsymbol{\beta}^{T} \mathbf{S} \boldsymbol{\beta}$ for a $K\times K$ matrix $\mathbf{S}$ \citep[Sec. 2]{woodInferenceComputationGeneralized2020}\footnote{For P-splines, $\mathbf{S}$ is a banded matrix not directly interpretable as based on derivatives, but \citet{woodPsplinesDerivativeBased2017} shows how to set up derivative based penalties also in this case.}. This gives a log likelihood penalizing deviations from linearity,
\begin{equation}
\label{eq:GAM_penalized_loglik}
    l\left(\boldsymbol{\beta}, \phi, \lambda\right) = \phi^{-1} \left(\mathbf{y}^{T} \mathbf{X} \boldsymbol{\beta} - b\left(\mathbf{X} \boldsymbol{\beta}\right)\right) + c\left(\mathbf{y},\phi\right)^{T} \mathbf{1}_{n} - \left(\lambda / 2\right)  \boldsymbol{\beta}^{T} \mathbf{S} \boldsymbol{\beta}.
\end{equation}
As shown by \citet{reissSmoothingParameterSelection2009} and \citet{woodFastStableRestricted2011}, estimation of $\lambda$ by maximizing either the restricted or marginal likelihood is less prone to overfitting in finite samples than prediction based criteria like generalized cross-validation \citep{golubGeneralizedCrossValidationMethod1979}. In this paper we use maximum marginal likelihood, and now illustrate how this allows interpreting \eqref{eq:GAM_penalized_loglik} as the log-likelihood of a GLMM, following \citet[Appendix]{woodStableEfficientMultiple2004}. 

First, form an eigendecomposition of the penalty matrix, $\mathbf{S} = \mathbf{U}\mathbf{D}\mathbf{U}^{T}$, yielding an orthogonal matrix $\mathbf{U} \in \mathbb{R}^{K\times K}$ and diagonal matrix $\mathbf{D} \in \mathbb{R}^{K\times K}$ with diagonal elements in decreasing order of magnitude. Let $\mathbf{D}^{+}$ be the $r \times r$ submatrix of $\mathbf{D}$ with nonzero entries on the diagonal, define $\boldsymbol{\beta}_{u} = \mathbf{U}^{T}\boldsymbol{\beta}$, and let $\mathbf{X}_{u} \in \mathbb{R}^{r}$ be the columns of $\mathbf{X}\mathbf{U}$ corresponding to $\mathbf{D}^{+}$ and $\mathbf{X}_{F} \in \mathbb{R}^{K -r}$ be the columns of $\mathbf{X}\mathbf{U}$ corresponding to zero entries on the diagonal of $\mathbf{D}$. Similarly, partition $\boldsymbol{\beta}_{u}$ into $\boldsymbol{\zeta}_{u} \in \mathbb{R}^{r}$ and $\boldsymbol{\beta}_{F} \in \mathbb{R}^{K-r}$ and let $\mathbf{X}_{R} = \mathbf{X}_{u} (\sqrt{\mathbf{D}^{+}})^{-1}$ and $\boldsymbol{\zeta} = \sqrt{\mathbf{D}^{+}} \boldsymbol{\zeta}_{u}$. We now have $\boldsymbol{\nu} = \mathbf{X}\boldsymbol{\beta} = \mathbf{X}_{F}\boldsymbol{\beta}_{F} + \mathbf{X}_{R} \boldsymbol{\zeta}$, and \eqref{eq:GAM_penalized_loglik} takes the form
\begin{equation}
\label{eq:GLM_mixed_penalized}
    l\left(\boldsymbol{\beta}_{F}, \boldsymbol{\zeta}, \phi, \lambda\right) = \phi^{-1}\left(\mathbf{y}^{T} \boldsymbol{\nu} - b\left(\boldsymbol{\nu}\right)\right) + c\left(\mathbf{y},\phi\right)^{T}\mathbf{1}_{n} - \left(\lambda / 2\right) \boldsymbol{\zeta}^{T} \boldsymbol{\zeta}.
\end{equation}
This is identical to the log-likelihood of a GLMM with fixed effects $\boldsymbol{\beta}_{F}$ of $\mathbf{X}_{F}$ and random effects $\boldsymbol{\zeta} \sim N(\mathbf{0}, \psi \mathbf{I})$ of $\mathbf{X}_{R}$, where $\psi = 1/\lambda$. The marginal likelihood is defined by integrating out the random effects from the joint density of $\mathbf{y}$ and $\boldsymbol{\zeta}$, which means computing the $r$-dimensional integral
\begin{equation}
\label{eq:GAM_marginal_likelihood}
    L\left(\boldsymbol{\beta}_{F}, \phi, \lambda\right) = 
     \left(2 \pi\right)^{-r/2}\int \exp\left(l\left(\boldsymbol{\beta}_{F}, \boldsymbol{\zeta}, \phi, \lambda\right)\right) \text{d}\boldsymbol{\zeta}.
\end{equation}
and then finding the values of $\boldsymbol{\beta}_{F}$, $\phi$, and $\lambda$ maximizing \eqref{eq:GAM_marginal_likelihood}. The Laplace approximation typically yields very good approximations to the integral \eqref{eq:GAM_marginal_likelihood} \citep[Sec. 2.2]{woodFastStableRestricted2011}. The final estimates $\hat{\boldsymbol{\zeta}}$ of $\boldsymbol{\zeta}$ would be taken as the modes of \eqref{eq:GLM_mixed_penalized} at the values $\hat{\boldsymbol{\beta}}_{F}$, $\hat{\phi}$, and $\hat{\lambda}$ maximizing \eqref{eq:GAM_marginal_likelihood}, and the spline weights in their original parametrization can be recovered by reverting the transformations leading up to \eqref{eq:GLM_mixed_penalized}. Imposing identifiability constraints on smooth terms requires an additional step in the above derivation, cf. \citet[Sec. 5.4.1]{woodGeneralizedAdditiveModels2017a}. 

$P$-values for smooth terms can be computed following \citet{woodPvaluesSmoothComponents2013} and approximate confidence bands following \citet{woodConfidenceIntervalsGeneralized2006} and \citet{marraCoveragePropertiesConfidence2012}. For the latter, let $\hat{\boldsymbol{\beta}}$ denote the estimated spline weights back in the original parametrization, and $\text{Cov}(\hat{\boldsymbol{\beta}}) \in \mathbb{R}^{K \times K}$ their covariance matrix. The estimates and squared standard errors at a new set of evaluation points $\mathbf{X}$ are now given by $\hat{\mathbf{f}} = \mathbf{X} \hat{\boldsymbol{\beta}}$ and $\hat{\mathbf{v}} = \text{diag}(\mathbf{X}\text{Cov}(\hat{\boldsymbol{\beta}}) \mathbf{X}^{T})$, and $(1 - \alpha)100$\% pointwise Wald type confidence bands are $\hat{\mathbf{f}} \pm z_{1 - \alpha/2} \sqrt{\hat{\mathbf{v}}}$, where $z_{1-\alpha/2}$ is the $\alpha/2$ quantile of the standard normal distribution \citep[Ch. 6.10]{woodGeneralizedAdditiveModels2017a}. Confidence bands constructed this way have close to nominal coverage averaged over the domain of the function \citep{marraCoveragePropertiesConfidence2012}. In contrast, simultaneous confidence bands covering the function over its whole domain with probability $(1-\alpha)100\%$ require a critical value $\tilde{z}_{\alpha/2}$ given by the $(1-\alpha)$th quantile of the random variable $r = \text{max}\{\mathbf{X}(\hat{\boldsymbol{\beta}} - \boldsymbol{\beta}) / \sqrt{\hat{\mathbf{v}}}\}$ \citep[Chapter 6.5]{ruppertSemiparametricRegression2003}. Since $\hat{\boldsymbol{\beta}} - \boldsymbol{\beta} \overset{\text{approx}}{\sim} N(\mathbf{0}, \text{Cov}(\hat{\boldsymbol{\beta}}))$ we can obtain an empirical Bayes posterior distribution of $r$ by simulation, and find $\tilde{z}_{\alpha/2}$ as the $(1-\alpha)$th quantile of $r$. A measure of the wiggliness of $\hat{\mathbf{f}}$ is given by its effective degrees of freedom, $(\mathbf{X}^{T}\mathbf{X} + \hat{\lambda} \mathbf{S})^{-1}\mathbf{X}^{T}\mathbf{X}$.

\subsection{Generalized linear latent and mixed models}

We now give a brief introduction to the GLLAMM framework for multilevel latent variable modeling, referring to \citet[Ch. 4.2-4.4]{skrondalGeneralizedLatentVariable2004} and \citet{rabe-heskethGeneralizedMultilevelStructural2004} for details. We still consider $n$ responses distributed according to \eqref{eq:galamm_distribution}, but now also assume that these elementary response units are clustered in $L$ levels. With $M_{l}$ latent variables at level $l$, the linear predictor for a single observational unit is \citep[eq. (4.9), p. 103]{skrondalGeneralizedLatentVariable2004}
\begin{equation}
\label{eq:GLLAMM_linpred}
    \nu = \boldsymbol{\beta}^{T} \mathbf{x} + \sum_{l=2}^{L}\sum_{m=1}^{M_{l}} \eta_{m}^{(l)} \boldsymbol{\lambda}_{m}^{(l)}{}^{T} \mathbf{z}_{m}^{(l)},
\end{equation}
omitting subscripts for observations. In \eqref{eq:GLLAMM_linpred}, $\mathbf{x}$ are explanatory variables with fixed effects $\boldsymbol{\beta}$, $\eta_{m}^{(l)}$ are latent variables varying at level $l$, and $\boldsymbol{\lambda}_{m}^{(l)}{}^{T} \mathbf{z}_{m}^{(l)}$ is the weighted sum of a vector of explanatory variables $\mathbf{z}_{m}^{(l)}$ varying at level $l$ and parameters $\boldsymbol{\lambda}_{m}^{(l)}$. Let $\boldsymbol{\eta}^{(l)} = [\eta_{1}^{(l)}, \dots, \eta_{M_{l}}^{(l)}]^{T} \in \mathbb{R}^{M_{l}}$ be the vector of all latent variables at level $l$, and $\boldsymbol{\eta} = [\boldsymbol{\eta}^{(2)}, \dots, \boldsymbol{\eta}^{(L)}]^{T} \in \mathbb{R}^{M}$ the vector of all latent variables belonging to a given level-2 unit, where $M = \sum_{l=2}^{L} M_{l}$. 

The structural model is given by
\begin{equation}
\label{eq:GLLAMM_structural}
    \boldsymbol{\eta} = \mathbf{B} \boldsymbol{\eta} + \boldsymbol{\Gamma} \mathbf{w} + \boldsymbol{\zeta},
\end{equation}
where $\mathbf{B}$ is an $M \times M$ matrix of regression coefficients for regression among latent variables and $\mathbf{w} \in \mathbb{R}^{Q}$ is a vector of $Q$ predictors for the latent variables with corresponding $M\times Q$ matrix of regression coefficients $\boldsymbol{\Gamma}$. $\boldsymbol{\zeta}$ is a vector of random effects, for which we use the same notation as defined for $\boldsymbol{\eta}$. Our framework is somewhat narrower than that of \citet{rabe-heskethGeneralizedMultilevelStructural2004, skrondalGeneralizedLatentVariable2004} as we assume normally distributed random effects, $\boldsymbol{\zeta}^{(l)} \sim N(\mathbf{0}, \boldsymbol{\Psi}^{(l)})$ for $l=2,\dots,L$, where $\boldsymbol{\Psi}^{(l)} \in \mathbb{R}^{M_{l} \times M_{l}}$ is the covariance matrix of random effects at level $l$. Defining the $M \times M$ covariance matrix $\boldsymbol{\Psi} = \text{diag}(\boldsymbol{\Psi}^{(2)}, \dots, \boldsymbol{\Psi}^{(L)})$, we also have $\boldsymbol{\zeta} \sim N(\mathbf{0}, \boldsymbol{\Psi})$. We assume recursive relations between latent variables, and require that a latent variable at level $l$ can only depend on latent variables varying at level $l$ or higher. It follows that $\mathbf{B}$ is strictly upper diagonal, if necessary after permuting the latent variables varying at each level \citep[p. 109]{rabe-heskethGeneralizedMultilevelStructural2004}.

Plugging the structural model \eqref{eq:GLLAMM_structural} into the linear predictor \eqref{eq:GLLAMM_linpred} yields the reduced form, which can then be inserted into the response model \eqref{eq:galamm_distribution} to give the joint density of $\mathbf{y}$ and $\boldsymbol{\zeta}$. Integrating $\boldsymbol{\zeta}$ out of this joint density gives the marginal likelihood. Proposed methods for maximizing this marginal likelihood include adaptive Gauss-Hermite quadrature integration combined with a Newton method \citep{rabe-heskethMaximumLikelihoodEstimation2005} and a profile likelihood algorithm based on Laplace approximation implemented in existing GLMM software \citep{jeonProfileLikelihoodApproachEstimating2012,rockwoodEstimatingComplexMeasurement2019}.

\section{Generalized additive latent and mixed models}
\label{sec:models}

We here present the proposed framework, which extends GLLAMMs to incorporate GAM-type nonlinear effects. Unless otherwise stated, the notation, basis functions, and distributional assumptions are as defined in Section \ref{sec:Background}.

\subsection{Proposed model framework}

We assume the response is distributed according to the exponential family \eqref{eq:galamm_distribution}, with the important extension that the functions $b(\cdot)$, $c(\cdot)$, and $g(\cdot)$ may vary between units, accomodating responses of mixed type. We modify the GLLAMM linear predictor \eqref{eq:GLLAMM_linpred} to
\begin{equation}
\label{eq:GALAMM_linpred}
\nu = \sum_{s=1}^{S} f_{s}\left(\mathbf{x}\right) + \sum_{l=2}^{L}\sum_{m=1}^{M_{l}} \eta_{m}^{(l)} \mathbf{z}^{(l)}_{m}{}^{'}\boldsymbol{\lambda}_{m}^{(l)},
\end{equation}
where $f_{s}(\mathbf{x})$, $s=1,\dots,S$ are smooth functions of a subset of explanatory variables $\mathbf{x}$. We also modify the structural part \eqref{eq:GLLAMM_structural} to allow the latent variables to depend smoothly on explanatory variables $\mathbf{w}$,
\begin{equation}
\label{eq:GALAMM_structural}
\boldsymbol{\eta} = \mathbf{B}\boldsymbol{\eta} + \mathbf{h}\left(\mathbf{w}\right)
+ \boldsymbol{\zeta},
\end{equation}
where $\mathbf{h}(\mathbf{w}) = [\mathbf{h}_{2}(\mathbf{w}), \dots, \mathbf{h}_{L}(\mathbf{w})] \in \mathbb{R}^{M}$ is a vector of smooth functions whose components $\mathbf{h}_{l}(\mathbf{w}) \in \mathbb{R}^{M_{l}}$ are vectors of functions predicting the latent variables varying at level $l$, and depending on a subset of the elements $\mathbf{w}$. We denote the scalar valued $m$th component of $\mathbf{h}_{l}(\mathbf{w})$ by $h_{lm}(\mathbf{w})$, and note that $h_{lm}(\mathbf{w})$ can only depend on elements of $\mathbf{w}$ varying at level $l$ or higher; otherwise the latent variable it predicts would vary at a level lower than $l$. If the $(l,m)$th latent variable is not predicted by any elements of $\mathbf{w}$, we set $h_{lm}(\mathbf{w}) = 0$. We assume that both $f_{s}(\mathbf{x})$ in \eqref{eq:GALAMM_linpred} and $h_{lm}(\mathbf{w})$ in \eqref{eq:GALAMM_structural} are smooth, as measured by their second derivatives. Together, the response distribution \eqref{eq:galamm_distribution}, linear predictor \eqref{eq:GALAMM_linpred}, and structural model \eqref{eq:GALAMM_structural} define a GALAMM with $L$ levels.

\subsection{Mixed model representation}
\label{sec:GALAMM_as_GLLAMM}

Using the mixed model representation of GAMs described in Section \ref{sec:GAM_intro}, we now show that an $L$-level GALAMM can be represented by an $(L+1)$-level GLLAMM, in which the $(L+1)$th level contains penalized spline coefficients.

First considering the linear predictor \eqref{eq:GALAMM_linpred}, we assume the $s$th smooth function $f_{s}(\mathbf{x})$ is represented by $K_{s}$ basis functions $b_{1,s}(\mathbf{x}), \dots, b_{K_{s},s}(\mathbf{x})$ with weights $\boldsymbol{\beta}_{s}$, yielding $S$ matrices $\mathbf{X}_{s} \in \mathbb{R}^{n \times K_{s}}$ with elements $(X_{s})_{ij} = b_{j,s}(\mathbf{x}_{i})$, for $s=1,\dots,S$, $j = 1, \dots, K_{s}$, and $i = 1,\dots, n$. Letting $\mathbf{f}_{s} \in \mathbb{R}^{n}$ denote the sample values of $f_{s}(\mathbf{x})$, we can repeat the steps leading up to \eqref{eq:GLM_mixed_penalized} to obtain $\mathbf{f}_{s} = \mathbf{X}_{s} \boldsymbol{\beta}_{s} = \mathbf{X}_{F,s} \boldsymbol{\beta}_{F,s} + \mathbf{X}_{R,s} \boldsymbol{\zeta}_{s}^{(L+1)}$, where $\boldsymbol{\zeta}_{s}^{(L+1)} \sim N(0, \boldsymbol{\Psi}_{s}^{(L+1)})$. Let $r_{s}$ denote the dimension of the range space of the smoothing matrix of $f_{s}(\mathbf{x})$, so $\boldsymbol{\Psi}_{s}^{(L+1)} \in \mathbb{R}^{r_{s} \times r_{s}}$, $\mathbf{X}_{R,s} \in \mathbb{R}^{n \times r_{s}}$, and $\mathbf{X}_{F,s} \in \mathbb{R}^{n \times (K_{s} - r_{s})}$. Repeating this for the $S$ smooth terms in the measurement model, we obtain the key terms defined in Table \ref{tab:linpred_key_terms}. The sample values of $\sum_{s=1}^{S} f_{s}(\mathbf{x})$ in the linear predictor \eqref{eq:GALAMM_linpred} are now given by $\sum_{s=1}^{S}\mathbf{f}_{s} = \mathbf{X}_{F} \boldsymbol{\beta}_{F} + \mathbf{X}_{R} \boldsymbol{\zeta}_{a}^{(L+1)}$, where $\boldsymbol{\zeta}_{a}^{(L+1)} \sim N(\mathbf{0}, \boldsymbol{\Psi}_{a}^{(L+1)})$.

\begin{table}
    \centering    
    \caption{Key terms in mixed effects representation of linear predictor \eqref{eq:GALAMM_linpred}.}
    \label{tab:linpred_key_terms}
    \begin{tabular}{ll}
    \toprule
         Description & Definition  \\
     \midrule
         Number of spline weights & $K = \sum_{s=1}^{S} K_{s}$ \\
         Number of random effects & $r_{a} = \sum_{s=1}^{S} r_{s}$ \\            
         Random effect predictors & $\mathbf{X}_{R} = [\mathbf{X}_{R,1}, \dots, \mathbf{X}_{R,S}] \in \mathbb{R}^{n \times r_{a}}$ \\
         Random effects & $\boldsymbol{\zeta}_{a}^{(L+1)} = [\boldsymbol{\zeta}_{1}^{(L+1)}{}^{T}, \dots, \boldsymbol{\zeta}_{S}^{(L+1)}{}^{T}]^{T} \in \mathbb{R}^{r_{a}}$ \\
         Random effects covariance & $\boldsymbol{\Psi}_{a}^{(L+1)} = \text{diag}(\boldsymbol{\Psi}_{1}^{(L+1)}, \dots, \boldsymbol{\Psi}_{S}^{(L+1)}) \in \mathbb{R}^{r_{a} \times r_{a}}$ \\
         Fixed effect predictors & $\mathbf{X}_{F} = [\mathbf{X}_{F,1}, \dots, \mathbf{X}_{F,S}] \in \mathbb{R}^{n \times (K - r_{a})}$ \\
         Fixed effects & $\boldsymbol{\beta}_{F} = [\boldsymbol{\beta}_{F,1}^{T}, \dots, \boldsymbol{\beta}_{F,S}^{T} ]^{T} \in \mathbb{R}^{K - r_{a}}$ \\         
    \bottomrule
    \end{tabular}      
\end{table}

Next considering the structural model \eqref{eq:GALAMM_structural}, we assume the $(l,m)$th smooth function is represented by $P_{lm}$ basis functions and define the matrix of sample values of basis functions as $\mathbf{W}_{lm} \in \mathbb{R}^{n_{2} \times P_{lm}}$ with elements $(W_{lm})_{ij} = b_{j,l,m}(\mathbf{w}_{i})$ for $j=1,\dots,P_{lm}$ and $i=1,\dots,n_{2}$, where $n_{2}$ is the total number of level-2 units. Eigendecomposing the smoothing matrix of $h_{lm}(\mathbf{w})$, the sample values can be written $\mathbf{h}_{lm} = \mathbf{W}_{lm} \boldsymbol{\gamma}_{lm} = \mathbf{W}_{F,lm} \boldsymbol{\gamma}_{F,lm} + \mathbf{W}_{R,lm} \boldsymbol{\zeta}_{lm}^{(L+1)} \in \mathbb{R}^{n_{2}}$, where $\mathbf{W}_{F,lm}$ contains the part of $h_{lm}(\mathbf{w})$ in the penalty nullspace, with fixed effects $\boldsymbol{\gamma}_{F,lm}$, and $\mathbf{W}_{R,lm} $ contains the components in the penalty range space, with random effects $\boldsymbol{\zeta}_{lm} \sim N(\mathbf{0}, \boldsymbol{\Psi}_{lm}^{(L+1)})$. Letting $r_{lm}$ denote the dimension of the range space of the smoothing matrix, we have $\boldsymbol{\Psi}_{lm}^{(L+1)} \in \mathbb{R}^{r_{lm} \times r_{lm}}$, $\mathbf{W}_{R,lm} \in \mathbb{R}^{n_{2} \times r_{lm}}$, and $\mathbf{W}_{F,lm} \in \mathbb{R}^{n_{2} \times (P_{lm} - r_{lm})}$. Repeating for all smooth functions predicting latent variables varying at level $l$, we obtain the level-$l$ terms in Table \ref{tab:structural_key_terms}, with $\boldsymbol{\zeta}_{l}^{(L+1)} \sim N(\mathbf{0}, \boldsymbol{\Psi}_{l}^{(L+1)})$. Next, repeating for smooth functions at all levels, we obtain the "all-level terms" in Table \ref{tab:structural_key_terms}, with $\boldsymbol{\zeta}_{b}^{(L+1)} \sim N(\mathbf{0}, \boldsymbol{\Psi}_{b}^{(L+1)})$.

\begin{table}
    \centering    
    \caption{Key terms in mixed effects representation of structural model \eqref{eq:GALAMM_structural}. In the bottom row, $\{\mathbf{e}_{1}, \dots , \mathbf{e}_{L-1} \}$ denotes the canonical basis for $\mathbb{R}^{L-1}$ and $\otimes$ is the Kronecker product.}
    \label{tab:structural_key_terms}
    \begin{tabular}{ll}
    \toprule
         Description & Definition  \\
         \midrule
         \multicolumn{2}{l}{\textit{Level-$l$ terms}} \\         
         \hspace{3mm} Random effect predictors & $\mathbf{W}_{R,l} = [\mathbf{W}_{R,l1}, \dots, \mathbf{W}_{R,lM_{l}}]$ \\
         \hspace{3mm} Random effects & $\boldsymbol{\zeta}_{l}^{(L+1)} = [\boldsymbol{\zeta}_{l1}^{(L+1)}{}^{T}, \dots, \boldsymbol{\zeta}_{lM_{l}}^{(L+1)}{}^{T}]^{T}$ \\
         \hspace{3mm} Random effects covariance & $\boldsymbol{\Psi}_{l}^{(L+1)} = \text{diag}(\boldsymbol{\Psi}_{l1}^{(L+1)}, \dots, \boldsymbol{\Psi}_{lM_{l}}^{(L+1)})$ \\
         \hspace{3mm} Fixed effect predictors & $\mathbf{W}_{F,l} = [\mathbf{W}_{F,l1}, \dots, \mathbf{W}_{F,lM_{l}}]$ \\
         \hspace{3mm} Fixed effects & $\boldsymbol{\gamma}_{F,l} = [\boldsymbol{\gamma}_{F,l1}^{T}, \dots, \boldsymbol{\gamma}_{F,lM_{l}}^{T}]^{T}$ \\    
         \midrule
         \multicolumn{2}{l}{\textit{All-level terms}} \\         
         \hspace{3mm} Number of spline weights & $P = \sum_{l=2}^{L} \sum_{m=1}^{M_{l}} P_{lm}$ \\
         \hspace{3mm} Number of random effects & $r_{b} = \sum_{l=2}^{L} \sum_{m=1}^{M_{l}} r_{lm}$ \\
         \hspace{3mm} Random effect predictors & $\mathbf{W}_{R} = [\mathbf{W}_{R,2}, \dots, \mathbf{W}_{R,L}] \in \mathbb{R}^{n_{2} \times r_{b}}$ \\
         \hspace{3mm} Random effects & $\boldsymbol{\zeta}_{b}^{(L+1)} = [\boldsymbol{\zeta}_{2}^{(L+1)}{}^{T}, \dots, \boldsymbol{\zeta}_{L}^{(L+1)}{}^{T} ]^{T} \in \mathbb{R}^{r_{b}}$ \\
         \hspace{3mm} Random effects covariance & $\boldsymbol{\Psi}_{b}^{(L+1)} = \text{diag}(\boldsymbol{\Psi}_{2}^{(L+1)}, \dots, \boldsymbol{\Psi}_{L}^{(L+1)}) \in \mathbb{R}^{r_{b} \times r_{b}}$ \\
         \hspace{3mm} Fixed effect predictors & $\mathbf{W}_{F} = [\mathbf{W}_{F,2}, \dots, \mathbf{W}_{F,L} ] \in \mathbb{R}^{n_{2} \times (P - r_{b})}$ \\
         \hspace{3mm} Fixed effects & $\mathbf{\Gamma} = [\mathbf{e}_{1} \otimes \boldsymbol{\gamma}_{F,2}^{T} , \dots, \mathbf{e}_{L-1} \otimes \boldsymbol{\gamma}_{F, L}^{T}] \in \mathbb{R}^{(L-1) \times (P-r_{b})}$ \\
    \bottomrule
    \end{tabular}    
\end{table}

Finally, we combine the random effects from the linear predictor summarized in Table \ref{tab:linpred_key_terms} and the structural model summarized in Table \ref{tab:structural_key_terms}, to get the vector of random effects at level $L+1$, $\boldsymbol{\zeta}^{(L+1)}   =
(\boldsymbol{\zeta}_{a}^{(L+1)}{}^{T},
     \boldsymbol{\zeta}_{b}^{(L+1)}{}^{T})^{T} \in \mathbb{R}^{M_{L+1}}$,
where $M_{L+1}=r_{a} + r_{b}$. It follows that $\boldsymbol{\zeta}^{(L+1)} \sim N(\mathbf{0}, \boldsymbol{\Psi}^{(L+1)})$ where $\boldsymbol{\Psi}^{(L+1)} = \text{diag}(\boldsymbol{\Psi}_{a}^{(L+1)}, \boldsymbol{\Psi}_{b}^{(L+1)}) \in \mathbb{R}^{M_{L+1} \times M_{L+1}}$. Let $\mathbf{x}_{F}^{T}$ and $\mathbf{x}_{R}^{T}$ correspond to rows of the matrices $\mathbf{X}_{F}$ and $\mathbf{X}_{R}$ defined in Table \ref{tab:linpred_key_terms}, i.e., the values for a single level-1 unit. Similarly let $\mathbf{w}_{F}^{T}$ and $\mathbf{w}_{R}^{T}$ correspond to rows of the matrices $\mathbf{W}_{F}$ and $\mathbf{W}_{R}$ defined in Table \ref{tab:structural_key_terms}, i.e., the values for a single level-2 unit. It follows that an $L$-level GALAMM with response \eqref{eq:galamm_distribution}, measurement model \eqref{eq:GALAMM_linpred}, and structural model \eqref{eq:GALAMM_structural} is identical to an $(L+1)$-level GLLAMM defined by
\begin{align}
\label{eq:GALAMM_mixed_linpred}
    & \nu = \boldsymbol{\beta}_{F}^{T} \mathbf{x}_{F} + \sum_{l=2}^{L+1} \sum_{m=1}^{M_{l}} \eta_{m}^{(l)} \mathbf{z}_{m}^{(l)}{}^{'}\boldsymbol{\lambda}_{m}^{(l)} \\
\label{eq:GALAMM_mixed_structural}
    & \boldsymbol{\eta} = \mathbf{B} \boldsymbol{\eta} + \boldsymbol{\Gamma} \mathbf{w}_{F} + \boldsymbol{\zeta},
\end{align}
where $\boldsymbol{\zeta} \sim N(\mathbf{0}, \boldsymbol{\Psi})$, with $\boldsymbol{\Psi} = \text{diag}(\boldsymbol{\Psi}^{(2)}, \dots, \boldsymbol{\Psi}^{(L+1)})$, subject to constraints which we now specify. Letting $x_{R,m}$ and $w_{R,m}$ denote the $m$th elements of $\mathbf{x}_{R}$ and $\mathbf{w}_{R}$, we require
\begin{align*}
& \mathbf{z}_{m}^{(L+1)} = 
\begin{cases}
    x_{R,m} & m = 1, \dots, r_{a} \\
    w_{R,m} \mathbf{z}_{n}^{(l)} & m = r_{a} + 1, \dots, M_{L+1}
\end{cases} \\
& \boldsymbol{\lambda}_{m}^{(L+1)} = 
\begin{cases}
    1 & m = 1, \dots, r_{a} \\
    \boldsymbol{\lambda}_{n}^{(l)} & m = r_{a} + 1, \dots, M_{L+1}
\end{cases}
\end{align*}
with $l$ in $\mathbf{z}_{n}^{(l)}$ and $\boldsymbol{\lambda}_{n}^{(l)}$, given $m$, defined by $l = \{l : \sum_{k=2}^{l-1} M_{k} < m \leq \sum_{k=2}^{l} M_{k}  m\}$, and given $l$ and $m$, $n = m - \sum_{k=2}^{l-1}M_{k}$. The first case in each constraint ensures that the random effects at level $L+1$ corresponding to smooth terms in the measurement model receive a factor loading equal to 1, and hence can be placed in the structural model. The second case in each constraint ensures that random effects at level $L+1$ corresponding to smooth terms predicting the $m$th latent variable at level $l$ are multiplied by the same factor loading and predictor as the fixed effect part of their smooth term when entering the linear predictor.

\section{Maximum marginal likelihood estimation}
\label{sec:Estimation}

We now present an algorithm for estimating both GALAMMs and GLLAMMs with normally distributed latent variables. An alternative approach would be to use the profile likelihood algorithm described by \citet{jeonProfileLikelihoodApproachEstimating2012,rockwoodEstimatingComplexMeasurement2019}, and we have confirmed that this algorithm gives practically identical estimates for the models considered in Section \ref{sec:latent_covariates} as well as simplified versions of the models considered in Section \ref{sec:latent_response}. However, for the applications considered in this paper, the proposed algorithm has been orders of magnitude faster, and it also offers increased flexibility by allowing mixed response types.

In the representation \eqref{eq:GALAMM_mixed_linpred}-\eqref{eq:GALAMM_mixed_structural}, the linear predictor for all $n$ elementary units of observation can be written $\boldsymbol{\nu} = \mathbf{X}(\boldsymbol{\lambda}, \mathbf{B}) \boldsymbol{\beta} +  \mathbf{Z}(\boldsymbol{\lambda}, \mathbf{B}) \boldsymbol{\zeta}$ \citep[eq. (4.21), p. 121]{skrondalGeneralizedLatentVariable2004}, where $\mathbf{X}(\boldsymbol{\lambda}, \mathbf{B}) \in \mathbb{R}^{n \times p}$ is a matrix of fixed effect predictors, with corresponding fixed effects $\boldsymbol{\beta} \in \mathbb{R}^{p}$, and $\mathbf{Z}(\boldsymbol{\lambda}, \mathbf{B}) \in \mathbb{R}^{n \times r}$ is a matrix of random effect predictors, with random effects $\boldsymbol{\zeta} \in \mathbb{R}^{r}$, $\boldsymbol{\zeta} \sim N(\mathbf{0}, \boldsymbol{\Psi})$. This notation makes it explicit that both matrices depend on factor loadings $\boldsymbol{\lambda}$ and regression coefficients between latent variables in $\mathbf{B}$. We allow dispersion parameters varying between observation by defining $\boldsymbol{\phi} \in \mathbb{R}^{n}$ with $i$th element $\phi_{g(i)}$, where $g(i)$ denotes the group $g$ to which the $i$th observation belongs. Following \citet{batesFittingLinearMixedEffects2015}, we write the covariance matrix in terms of a relative covariance factor $\boldsymbol{\Lambda} \in \mathbb{R}^{r \times r}$, $\boldsymbol{\Psi} = \phi_{1}\boldsymbol{\Lambda} \boldsymbol{\Lambda}^{T}$, where the dispersion parameter for group 1 is used as reference level. 

The matrices $\mathbf{Z}(\boldsymbol{\lambda}, \mathbf{B})$ and $\boldsymbol{\Lambda}$ are often very sparse, and sparse matrix methods have been shown to be efficient in the case of LMMs \citep{batesFittingLinearMixedEffects2015,fraleyLargeScaleEstimationVariance1995}. With nested random effects, algorithms using dense matrix methods can also be efficient \citep{pinheiroApproximationsLogLikelihoodFunction1995,pinheiroMixedEffectsModelsSPLUS2000, pinheiroEfficientLaplacianAdaptive2006,rabe-heskethMaximumLikelihoodEstimation2005}, but these methods scale poorly with crossed random effects. The R package \verb!lme4! uses sparse matrix methods also for GLMMs and nonlinear mixed models with normally distributed responses, as described in a package vignette \citep{batesComputationalMethodsMixed2022}. We here extend these methods to the case of GALAMMs, the key differences being the presence of mixed response types, the parameters $\boldsymbol{\lambda}$ and $\mathbf{B}$, and our use of automatic differentiation to obtain derivatives of the marginal likelihood to machine precision. We assume throughout that necessary identifiability constraints have been imposed.

\subsection{Evaluating the marginal likelihood}
\label{sec:evaluating_likelihood}

Through the transformation $\boldsymbol{\Lambda} \mathbf{u} = \boldsymbol{\zeta}$, we define uncorrelated random effects $\mathbf{u} \in \mathbb{R}^{r}$ distributed according to $N(\mathbf{0}, \phi_{1} \mathbf{I}_{r})$ \citep{batesFittingLinearMixedEffects2015}. Integrating over these random effects yields the marginal likelihood
\begin{equation}
\label{eq:GLLAMM_marginal_likelihood}
    L\left(\boldsymbol{\beta}, \boldsymbol{\Lambda}, \boldsymbol{\Gamma}, \boldsymbol{\lambda}, \mathbf{B}, \boldsymbol{\phi}\right) =  \left(2 \pi \phi_{1}\right)^{-r/2}  \int_{\mathbb{R}^{r}} \exp\left( g\left(\boldsymbol{\beta}, \boldsymbol{\Lambda}, \boldsymbol{\Gamma}, \boldsymbol{\lambda}, \mathbf{B}, \boldsymbol{\phi}, \mathbf{u}\right) \right) \text{d} \mathbf{u},
\end{equation}
with the term in the exponent given by
\begin{equation}
\label{eq:GLLAMM_integrand}
    g\left(\boldsymbol{\beta}, \boldsymbol{\Lambda}, \boldsymbol{\Gamma}, \boldsymbol{\lambda}, \mathbf{B}, \boldsymbol{\phi}, \mathbf{u}\right) = \mathbf{y}^{T} \mathbf{W}\boldsymbol{\nu} - d\left(\boldsymbol{\nu}\right)^{T} \mathbf{W}\mathbf{1}_{n}  + c\left(\mathbf{y}, \boldsymbol{\phi}\right)^{T} \mathbf{1}_{n} - \left(2\phi_{1}\right)^{-1} \left\| \mathbf{u} \right\|^{2},
\end{equation}
where $\mathbf{W} =  \text{diag}\{\boldsymbol{\phi}^{-1}\} \in \mathbb{R}^{n\times n}$ and we omit in the notation that $c(\cdot)$ and $d(\cdot)$ may vary between observations. Define the conditional modes of $\mathbf{u}$ as
\begin{equation}
\label{eq:conditional_modes}
    \tilde{\mathbf{u}} = \underset{\mathbf{u}}{\text{argmax}} \left\{ g\left(\boldsymbol{\beta}, \boldsymbol{\Lambda}, \boldsymbol{\Gamma}, \boldsymbol{\lambda}, \mathbf{B}, \phi, \mathbf{u}\right) \right\}.
\end{equation}
Following \citet{pinheiroEfficientLaplacianAdaptive2006}, these modes can be found with penalized iteratively reweighted least squares, by noting that the gradient and Hessian of $g(\cdot)$ with respect to $\mathbf{u}$ are
\begin{align*}
 & \boldsymbol{\nabla} g = \boldsymbol{\Lambda}^{T} \mathbf{Z}^{T} \mathbf{W}\left( \mathbf{y} - \boldsymbol{\mu} \right)  - \left(1/\phi_{1}\right)\mathbf{u} \in \mathbb{R}^{r} \\
 & \mathbf{H}_{g} =-  \boldsymbol{\Lambda}^{T} \mathbf{Z}^{T} \mathbf{V}  \mathbf{Z} \boldsymbol{\Lambda} - \left(1/\phi_{1}\right) \mathbf{I}_{r}   \in \mathbb{R}^{r \times r},
\end{align*}
where $\boldsymbol{\mu} = d'(\boldsymbol{\nu})$ and $\mathbf{V} \in \mathbb{R}^{n \times n}$ is a diagonal matrix with $i$th diagonal element $d''(\nu_{i}) /\phi_{g(i)}$. 

We form a sparse Cholesky factorization \citep{davisDirectMethodsSparse2006} of the Hessian, $\mathbf{L} \mathbf{D} \mathbf{L}^{T} = -\mathbf{P}  \mathbf{H}_{g} \mathbf{P}^{T}$, where $\mathbf{L}  \in \mathbb{R}^{r \times r}$ is lower triangular, $\mathbf{D} \in \mathbb{R}^{r \times r}$ is diagonal, and $\mathbf{P}\in \mathbb{R}^{r \times r}$ is a permutation matrix chosen to minimize the number of operations in the Gaussian elimination steps for solving a linear system of the form $\mathbf{L} \mathbf{D} \mathbf{L}^{T} \mathbf{x} = \mathbf{b}$, as we do in \eqref{eq:PIRLS} below. Importantly, $\mathbf{P}$ only depends on the location of the structural zeroes, and not on particular values of the nonzero elements of the Hessian. $\mathbf{P}$ can hence be computed a single time for some initial values of the parameters, and then stored for reuse in all subsequent iterations. We used the approximate minimum degree algorithm of \citet{amestoyApproximateMinimumDegree1996} for defining $\mathbf{P}$, which is further described in \citet[Ch. 7]{davisDirectMethodsSparse2006} and \citet[Ch. 11.3]{duffDirectMethodsSparse2017}.

A Newton method for finding the conditional modes \eqref{eq:conditional_modes} starts at an initial estimate $\mathbf{u}^{(0)}$ and then at step $k$ solves the linear system $\mathbf{H}_{g}^{(k)} \boldsymbol{\delta}^{(k)} =\boldsymbol{\nabla} g^{(k)}$, whereupon the estimates are updated with $\mathbf{u}^{(k+1)} = \mathbf{u}^{(k)} + \tau \boldsymbol{\delta}^{(k)}$ for some stepsize $\tau$ ensuring that $g(\cdot)$ increases at each step \citep[eq. 40]{batesComputationalMethodsMixed2022}. In terms of the sparse matrix representation, at each iteration the Cholesky factorization must first be updated so it satisfies $\mathbf{L}^{(k)} \mathbf{D}^{(k)} \mathbf{L}^{(k)}{}^{T} = -\mathbf{P}  \mathbf{H}_{g}^{(k)} \mathbf{P}^{T}$ and then the linear system 
\begin{equation}
\label{eq:PIRLS}
 \mathbf{L}^{(k)} \mathbf{D}^{(k)} \mathbf{L}^{(k)T} \mathbf{P} \boldsymbol{\delta}^{(k)} = \mathbf{P} \left( \boldsymbol{\Lambda}^{T} \mathbf{Z}^{T} \mathbf{W}^{(k)} \left( \mathbf{y} - \boldsymbol{\mu}^{(k)}\right)- (1/\phi_{1}^{(k)}) \mathbf{u}^{(k)}\right).
\end{equation}
must be solved for $\boldsymbol{\delta}^{(k)}$. The superscript in $\mathbf{W}^{(k)}$ is due to the fact that for some distributions, e.g., the normal, the explicit formula for the dispersion parameter depends on $\mathbf{u}$. Our implementation uses step-halving, i.e., starting from $\tau = 1$, $\tau \leftarrow \tau / 2$ is repeated until $g(\boldsymbol{\beta}, \boldsymbol{\Lambda}, \boldsymbol{\Gamma}, \boldsymbol{\lambda}, \mathbf{B}, \boldsymbol{\phi}, \mathbf{u}^{(k+1)}) > g(\boldsymbol{\beta}, \boldsymbol{\Lambda}, \boldsymbol{\Gamma}, \boldsymbol{\lambda}, \mathbf{B}, \boldsymbol{\phi}, \mathbf{u}^{(k)})$. In the case of Gaussian responses and unit link function, \eqref{eq:PIRLS} is solved exactly in a single step. 

At convergence at some $k$, we set $\tilde{\mathbf{u}} = \mathbf{u}^{(k)}$, $\mathbf{L} = \mathbf{L}^{(k)}$, and $\mathbf{D} = \mathbf{D}^{(k)} $. A second order Taylor expansion of \eqref{eq:GLLAMM_integrand} around its mode is then given by
\begin{equation}
\label{eq:GLLAMM_integrand_taylor}
    g\left(\boldsymbol{\beta}, \boldsymbol{\Lambda}, \boldsymbol{\Gamma}, \boldsymbol{\lambda}, \mathbf{B}, \boldsymbol{\phi}, \mathbf{u}\right)  \approx g\left(\boldsymbol{\beta}, \boldsymbol{\Lambda}, \boldsymbol{\Gamma}, \boldsymbol{\lambda}, \mathbf{B}, \boldsymbol{\phi}, \tilde{\mathbf{u}}\right) - (1/2)\left(\mathbf{u} -\tilde{\mathbf{u}} \right)^{T} \mathbf{P}^{T} \mathbf{L} \mathbf{D} \mathbf{L}^{T} \mathbf{P} \left(\mathbf{u} -\tilde{\mathbf{u}} \right).
\end{equation}
The Laplace approximation uses \eqref{eq:GLLAMM_integrand_taylor} to approximate the marginal likelihood \eqref{eq:GLLAMM_marginal_likelihood} with
\begin{align*}
 L(\boldsymbol{\beta}, \boldsymbol{\Lambda}, \boldsymbol{\Gamma}, \boldsymbol{\lambda}, \mathbf{B}, \boldsymbol{\phi}) \approx 
  \exp(g(\boldsymbol{\beta}, \boldsymbol{\Lambda}, \boldsymbol{\Gamma}, \boldsymbol{\lambda}, \mathbf{B}, \boldsymbol{\phi}, \tilde{\mathbf{u}})) | \mathbf{P}^{T} \mathbf{L} \sqrt{\mathbf{D}}|^{-1}.
\end{align*}
It follows that the Laplace approximate marginal log-likelihood is
\begin{align}
    \label{eq:GLLAMM_laplace_loglik}
    &\log L\left(\boldsymbol{\beta}, \boldsymbol{\Lambda}, \boldsymbol{\Gamma}, \boldsymbol{\lambda}, \mathbf{B}, \boldsymbol{\phi}\right) = \\ \nonumber
    &\mathbf{y}^{T} \mathbf{W}\boldsymbol{\nu} - d\left(\boldsymbol{\nu}\right)^{T} \mathbf{W}\mathbf{1}_{n}  + c\left(\mathbf{y}, \boldsymbol{\phi}\right)^{T} \mathbf{1}_{n} - \left(2\phi_{1}\right)^{-1} \left\| \tilde{\mathbf{u} }\right\|^{2} - (1/2)\log \text{tr}\left(\mathbf{D}\right),
\end{align}
where all terms are evaluated at $\tilde{\mathbf{u}}$ and we used the identity $\log |\mathbf{P}^{T} \mathbf{L} \sqrt{\mathbf{D}}|^{-1} = -(1/2) \log \text{tr} (\mathbf{D})$, $\text{tr}(\cdot)$ denoting matrix trace.

\subsection{Maximizing the marginal likelihood}
\label{sec:maximum_likelihood}

Having an iterative algorithm for computing the Laplace approximate marginal log-likelihood \eqref{eq:GLLAMM_laplace_loglik}, we now consider the problem of maximizing it. This is a constrained optimization problem since, e.g., elements of $\boldsymbol{\Lambda}$ and $\boldsymbol{\phi}$ may be required to be non-negative. We here treat the general problem of maximizing the marginal likelihood with respect to all its parameters, but note that in special cases the dimension of the optimization problem can be reduced. For example, in the Gaussian unit link case, expressions for values of $\boldsymbol{\beta}$ and $\boldsymbol{\phi}$ maximizing \eqref{eq:GLLAMM_laplace_loglik} given the other parameters are directly available \citep[Sec. 3.4]{batesFittingLinearMixedEffects2015}. 

For each new set of candidate parameters, the terms in \eqref{eq:GLLAMM_laplace_loglik} also need to be updated, and for $\mathbf{X}$, $\mathbf{Z}$, and $\boldsymbol{\Lambda}$ this requires special care. For $\boldsymbol{\Lambda}$, we use the mapping between the structural non-zeros of $\boldsymbol{\Lambda}$ and fundamental parameters described in \citet[pp. 11-13]{batesFittingLinearMixedEffects2015}. Updating of $\mathbf{Z}$ was obtained by initializing $\mathbf{Z}$ with $\boldsymbol{\lambda}$ and $\mathbf{B}$ set at some default values, and a function $f_{i}(\boldsymbol{\lambda}, \mathbf{B})$ representing a factor the $i$th structural nonzero of $\mathbf{Z}$ needs to be multiplied with. Hence, if $z_{i}$ denotes the $i$th structural nonzero of $\mathbf{Z}$, it gets updated according to $z_{i} \leftarrow z_{i} \times f_{i}(\boldsymbol{\lambda}, \mathbf{B})$. An equivalent approach was used for $\mathbf{X} \in \mathbb{R}^{n \times p}$, but since this matrix typically is dense, with the number of fixed effects $p$ being relatively low, the updating iteration was performed over all matrix elements.

Forward mode automatic differentiation with first-order dual numbers was used to evaluate the gradient of \eqref{eq:GLLAMM_laplace_loglik} with respect to all its parameters, by extending the sparse matrix methods provided by the \verb!C++! library \verb!Eigen! \citep{guennebaudEigenV32010} with dual numbers provided by the \verb!C++! library \verb!autodiff! \citep{lealAutodiffModernFast2018}, using template metaprogramming \citep{meyersEffectiveModern2015}. Automatic differentiation exploits the fact that every computer program performs a set of elementary arithmetic operations, so by repeatedly applying the chain rule derivatives are obtained with accuracy at machine precision \citep{baydinAutomaticDifferentiationMachine2018,margossianReviewAutomaticDifferentiation2019,skaugAutomaticDifferentiationFacilitate2002}. Next, the gradients were used by the L-BFGS-B algorithm \citep{byrdLimitedMemoryAlgorithm1995} implemented in \verb!R!'s \verb!optim()! function \citep{rcoreteamLanguageEnvironmentStatistical2022} to maximize the log-likelihood. L-BFGS-B is a quasi-Newton method which uses gradient information to approximate the Hessian matrix and gradient projection to keep the solutions inside the feasible set \citep[Ch. 7.2]{nocedalNumericalOptimization2006}. \verb!RcppEigen! \citep{batesFastElegantNumerical2013} was used for interfacing \verb!R! and \verb!C++!, and the \verb!memoise! package \citep{wickhamMemoiseMemoisationFunctions2021} for caching during optimization. 

At convergence, the Hessian of \eqref{eq:GLLAMM_laplace_loglik} with respect to parameters of interest can be computed using forward mode automatic differentiation with second-order dual numbers. The negative inverse of this matrix is the asymptotic covariance matrix, which can then be used to compute Wald type confidence intervals for parameters and pointwise and simultaneous confidence bands for smooth terms, as described in the last paragraph of Section \ref{sec:GAM_intro}. A requirement for such uncertainty estimation to work well is that the marginal log-likelihood \eqref{eq:GLLAMM_laplace_loglik} is well approximated by a quadratic function in a region near its maximum, i.e., that we are sufficiently close to the asymptotic regime \citep[Ch. 5.2-5.3]{pawitanAllLikelihood2001}. In Section \ref{sec:cog_sim} below we describe a parametric bootstrapping procedure which can be used to check this assumption.

\section{Latent response model with factor-by-curve interaction and mixed response types}
\label{sec:latent_response}

\subsection{Estimating lifespan trajectories of abilities in three cognitive domains}
\label{sec:cog_app}

Dating back at least to \citet{spearmanGeneralIntelligenceObjectively1904}, individual abilities in cognitive domains are known to be correlated, and a recent meta analysis has confirmed that also change in cognitive abilities during adulthood is highly correlated across domains \citep{tucker-drobCoupledCognitiveChanges2019}. However, a topic which has been more debated is the timing of age-related decline in cognitive function \citep{nilssonChallengingNotionEarlyonset2009,razOnlyTimeWill2011,salthouseNeuroanatomicalSubstratesAgerelated2011,schaieWhenDoesAgerelated2009}, with cross-sectional studies indicating that the decline starts around the age of 20 \citep{salthouseWhenDoesAgerelated2009} and longitudinal studies showing a stable level until the age of 60 \citep{ronnlundStabilityGrowthDecline2005}. Furthermore, cognitive abilities involving fluid reasoning typically peak earlier than crystallized knowledge, which depends more on previously acquired knowledge \citep[Fig. 1]{tucker-drobCognitiveAgingDementia2019}. Common to all the mentioned studies is the use of purely parametric models, typically linear, or categorization into discrete age groups which have been analyzed separately.

In this section we demonstrate how GALAMMs can be used to estimate lifespan trajectories of abilities in three cognitive domains involving fluid reasoning, using data from the Center for Lifespan Changes in Brain and Cognition \citep{fjellNeuroinflammationTauInteract2018,walhovdNeurodevelopmentalOriginsLifespan2016}. \emph{Episodic memory} involves recollection of specific events, for which the California verbal learning test (CVLT) \citep{delisCVLTCaliforniaVerbal1987,delisCVLTCaliforniaVerbal2000} is widely used. During the test, the experimenter reads a list of 16 words aloud, and subsequently the participant is asked to repeat the words back. The procedure is repeated in five trials, as well as two delayed trials after 5 and 30 minutes. Each complete CVLT hence gives 7 elementary units of observation recording the number of successes in 16 trials. \emph{Working memory} involves the ability to hold information temporarily and can be assessed by digit span tests, in which a sequence of numbers of increasing length is read out loud, and the participant is asked to immediately repeat the digits back \citep{blackburnRevisedAdministrationScoring1959,ostrosky-solisDigitSpanEffect2006}. The initial list was of length 2, step-wise increasing to length 9, and then repeated once more. The final score was an integer between 0 and 16 representing the total number of lists correctly recalled. The data also contained results from an otherwise identical digit span backwards task \citep{hilbertDigitSpanBackwards2015}, in which the participants were asked to repeat the list of numbers backwards. Hence, each digit span test contained at least two elementary units of observation, one for the forward task and one for the backward task. The Stroop test is a test of \emph{executive function} and \emph{processing speed}\footnote{For simplicity we use the term 'executive function' in what follows.} \citep{scarpinaStroopColorWord2017,siscoParkinsonDiseaseStroop2016,stroopStudiesInterferenceSerial1935}. The D-KEFS version \citep{delisDelisKaplanExecutiveFunction2001} was used,  consisting of four tests \citep[p. 797]{fineDelisKaplanExecutive2011}. Baseline conditions 1 and 2 involve naming of color squares and reading of color words printed in black. In condition 3 color names are printed in ink which conflicts with the color name, and the participant must name the color (e.g., if the word 'blue' is printed in red, the participant must read 'red'), the point being that to persons who can read, reading is more automatic than retrieving color names, so there is a conflict. In condition 4, the participant must switch between naming colors as in condition 3 and reading words printed in dissonant ink color. Each of the four conditions constitutes an elementary unit of observation, and each response is a measure of the time taken to complete the tests under the condition. 

\begin{figure}
\centering
\includegraphics[width=\columnwidth]{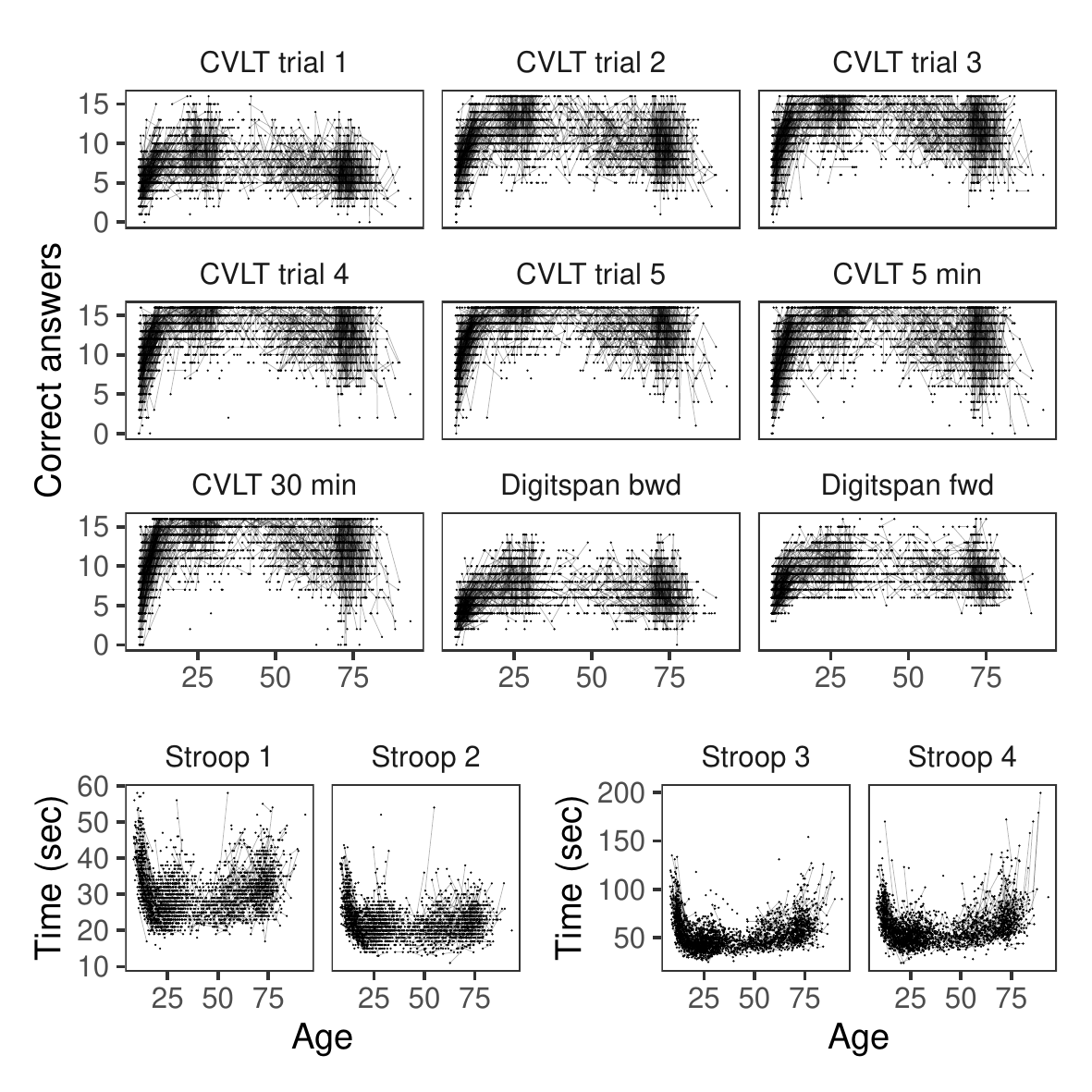}
\caption{\textbf{Cognitive test scores.} Observed responses to the thirteen test scores used in Section \ref{sec:cog_app}, plotted versus age. Dots show individual responses, and gray lines connect multiple timepoints for the same participant.}
\label{fig:cognition_data}
\end{figure}

The CVLT trials consisted of 24147 observations of 1873 healthy individuals, the digit span trials of 6758 observations from 1858 individuals, and the Stroop trials of 9929 observations from 1695 individuals, with a large degree of overlap between tests. In total, there were 40834 elementary units of observation, the number of timepoints for each individual varied between 1 and 6, and the time interval between two consecutive measurements varied between 11 days and 9.9 years, with mean interval 2.4 years. Further details about the data can be found in Online Resource 1. 

Figure \ref{fig:cognition_data} shows plots of the observed responses, illustrating that the scores on each test vary nonlinearly across the lifespan\cprotect\footnote{Figure \ref{fig:cognition_data} and all subsequent plots were created in R using \verb!ggplot2! \citep{wickhamGgplot2ElegantGraphics2016}, \verb!patchwork! \citep{pedersenPatchworkComposerPlots2020}, \verb!ggthtemes! \citep{arnoldGgthemesExtraThemes2021}, and \verb!gghalves! \citep{tiedemannGghalvesComposeHalfhalf2020}}. For CVLT we see that the participants recalled a larger number of words in later trials, illustrating a within-timepoint learning effect. Ceiling effects were also apparent in later CVLT trials, as a large number of participants remembered all 16 words. For the digit span tests, it is clear that the backward test is more challenging than the forward test, as illustrated by the lower number of correct answers. For the Stroop test, the relationship with the latent ability is inverted, as a low time to completion implies high performance. The higher times to completion for conditions 3 and 4 in the Stroop test show that these are more challenging than conditions 1 and 2.

Assuming that the number of correct answers to the CVLT tests are noisy measurements of the participants' episodic memory, that the number of correct answers to the digit span tests are noisy measurements of working memory, and that the negative of the time required to complete the Stroop tests are noisy measurements of executive function, our goal was to estimate how abilities in these domains vary with age. We defined a three-level GALAMM, in which the first level contained the elementary units of observation, the second level contained all tests taken by an individual at a given timepoint, and the third level contained each individual participant. For CVLT and digit span tests, the responses $y_{i} \in \{0,\dots, 16\}$ were assumed binomially distributed, using a logit link $\nu_{i} = g(\mu_{i}) = \log (\mu_{i} / (1 - \mu_{i}))$, where $\mu_{i}$ was the expected proportion of successes. For the continuous responses from the Stroop tests, a normal distribution with unit link function was used. 

The measurement model took the form\cprotect\footnote{Because an inner product is computed between both $\boldsymbol{\beta}_{t}$ and $\boldsymbol{\lambda}_{m}$ and the variable $\mathbf{z}_{ti}$, we use the letter $\mathbf{z}$ for all terms in the measurement model, although the letter $\mathbf{x}$ would be more consistent with the definition \eqref{eq:GALAMM_linpred}.}
\begin{equation}
\label{eq:cognitive_example_measurement}
 \nu_{i} = \mathbf{z}_{ti}^{T} \boldsymbol{\beta}_{t} + \mathbf{z}_{ri}^{T} \boldsymbol{\beta}_{r} + \sum_{m=1}^{3} \mathbf{z}_{ti}^{T} \boldsymbol{\lambda}_{m} \eta_{m},
\end{equation}
where $\mathbf{z}_{ti}$ is an indicator vector of size 13 whose $k$th element equals one if the $i$th elementary unit of observation is the $k$th test in the order of appearance in Figure \ref{fig:cognition_data}. Accordingly, $\boldsymbol{\beta}_{t} \in \mathbb{R}^{13}$ was a vector of trial effects. Retest effects, which can be defined as the marginal effect of having taken the test previously, have been documented for all the three tests used in this study \citep{davidsonStroopInterferencePractice2003,steeleListeningMozartDoes1997,woodsCaliforniaVerbalLearning2006} and were accounted for by the term $\mathbf{z}_{ri}^{T}\boldsymbol{\beta}_{r}$. Due to the different scales of the responses in Stroop conditions 1 and 2 compared to conditions 3 and 4, both retest effects and residual standard errors were estimated independently for these two groups. Accordingly, $\mathbf{z}_{ri}$ was a vector of size 4, whose first element was an indicator for the event that the participant had taken the CVLT at a previous time, the second element a corresponding indicator for the digit span test, the third element for Stroop condition 1 or 2, and the fourth element for Stroop condition 3 or 4. Thus, $\boldsymbol{\beta}_{r} = (\beta_{r1}, \beta_{r2}, \beta_{r3}, \beta_{r4})^{T}$ contained retest effects for CVLT, digit span, Stroop conditions 1 and 2, and Stroop conditions 3 and 4. Considering the last term in \eqref{eq:cognitive_example_measurement}, $\boldsymbol{\lambda}_{1} \in \mathbb{R}^{7}$ contained loadings relating the CVLT trials to latent episodic memory $\eta_{1}$, $\boldsymbol{\lambda}_{2} \in \mathbb{R}^{2}$ contained loadings relating digit span scores to latent working memory $\eta_{2}$, and $\boldsymbol{\lambda}_{3}$ contained loadings relating Stroop scores to latent executive function $\eta_{3}$. In $\boldsymbol{\lambda}_{1}$ and $\boldsymbol{\lambda}_{2}$, the first element was constrained to 1 for identifiability, and in $\boldsymbol{\lambda}_{3}$ it was constrained to $-1$, since a high time taken in each Stroop condition is associated with lower executive function. During model estimation, the results for Stroop conditions 1 and 2 and Stroop condition 3 and 4 were standardized to have zero mean and unit variance, but the results shown are transformed back to units of seconds.

Next, we used the structural model
\begin{equation}
    \label{eq:cognitive_example_structural}
    \eta_{m} = h_{m}\left(w\right) + \zeta_{m}^{(2)} + \zeta_{m}^{(3)}, ~ m=1,2,3,
\end{equation}
where $w$ denotes age. The smooth functions $h_{m}(w)$ model the lifespan trajectories of abilities, with $m=1$ denoting episodic memory, $m=2$ working memory, and $m=3$ executive function. The level-2 random intercepts $\zeta_{m}^{(2)} \sim N(0, \psi_{m}^{(2)})$, varying between timepoints for the same participant were assumed uncorrelated, taking the role of residuals in the structural model \eqref{eq:cognitive_example_structural}. Level-3 random intercepts $\boldsymbol{\zeta}^{(3)} = (\zeta_{1}^{(3)}, \zeta_{2}^{(3)}, \zeta_{3}^{(3)})' \sim N(\mathbf{0}, \boldsymbol{\Psi}^{(3)})$ had a freely estimated covariance matrix $\boldsymbol{\Psi}^{(3)} \in \mathbb{R}^{3\times 3}$ with six non-redundant parameters. Each smooth term $h_{m}(w)$ was constructed from 10 cubic regression splines subject to sum-to-zero constraints \citep[Ch. 5.4.1]{woodGeneralizedAdditiveModels2017a}, and had its own smoothing parameter. Estimating the model using the algorithm described in Section \ref{sec:Estimation} took about five hours, and the proportion of structural zeroes in the random effects design matrix $\mathbf{Z}$ used in model fitting was 99.9\%.

The estimated regression coefficients and factor loadings are summarized in Table \ref{tab:cognition_parametric_terms}. Considering episodic memory first, the trial effects increased with trial number 1-5, reflecting that participants on average achieved higher scores in later trials, while the effects declined again in the delayed trials, indicating increasing difficulty. The factor loadings for CVLT were markedly higher in the later trials, indicating that these trials have a better ability to discriminate between high and low values of latent episodic memory. For digit span tests, the factor loadings were of similar magnitude, indicating that the tests had similar ability to discriminate between latent working memory, but as expected the trial effect for the forward test was higher than the backward test, since it is easier. For the Stroop tests, the factor loadings for the time taken to complete the trials under condition 3 and 4 were considerably higher than for conditions 1 and 2, reflecting the increased variance for these more challenging trials\cprotect\footnote{The factor loading for Stroop condition 1 was fixed to $-1$ on the scale used during fitting. Transformed back to the original parametrization it took to the value $-7.05$ seconds.}. As expected, significant retest effects were found for each test. For Stroop, having taken the test previously was associated with 1.24 seconds lower time to completion under conditions 1 and 2, and 2.21 seconds lower time under conditions 3 and 4. We also note that simulation experiments reported in Section \ref{sec:cog_sim} suggest that confidence intervals for the factor loadings for Stroop conditions 2-4 and the trial effects for Stroop conditions 1 and 2 should be based on bootstrapping rather than using the Wald procedure with the asymptotic standard error reported in Table \ref{tab:cognition_parametric_terms}. Complete bootstrap and Wald type confidence intervals for all parameters in Table \ref{tab:cognition_parametric_terms} are given in Tables S2 and S3 of Online Resource 1. 

\begin{table}
\centering
\begin{threeparttable}
\caption{Estimates and standard errors of parametric terms in the model presented in Section \ref{sec:cog_app}.}
\label{tab:cognition_parametric_terms}
\small
\begin{tabular}{lll}
\toprule
Parameter & Trial effect (SE) & Factor loading (SE) \\
\midrule
\multicolumn{3}{l}{\textit{Episodic memory}}\\
\hspace{3mm} CVLT trial 1 & $\beta_{t1} = -0.26$ (0.01)  & $\lambda_{11} = 1$ (-) \\
\hspace{3mm} CVLT trial 2 & $\beta_{t2} = 0.74$ (0.02)  & $\lambda_{12} = 1.79$ (0.03) \\
\hspace{3mm} CVLT trial 3 & $\beta_{t3} = 1.42$ (0.02)  & $\lambda_{13} = 2.44$ (0.05) \\
\hspace{3mm} CVLT trial 4 & $\beta_{t4} = 1.84$ (0.03)  & $\lambda_{14} = 2.76$ (0.05) \\
\hspace{3mm} CVLT trial 5 & $\beta_{t5} = 2.17$ (0.03)  & $\lambda_{15} = 3.02$ (0.06) \\
\hspace{3mm} CVLT 5 min delay & $\beta_{t6} = 1.62$ (0.03)  & $\lambda_{16} = 3.04$ (0.06) \\
\hspace{3mm} CVLT 30 min delay & $\beta_{t7} = 1.81$ (0.03)  & $\lambda_{17} = 3.22$ (0.06) \\
\midrule
\multicolumn{3}{l}{\textit{Working memory}} \\
\hspace{3mm} Digit span backward & $\beta_{t8} = -0.39$ (0.01) & $\lambda_{21} = 1$  (-) \\
\hspace{3mm} Digit span forward & $\beta_{t9} = 0.29$ (0.01) & $\lambda_{22} = 0.96$  (0.03)\\
\midrule
\multicolumn{3}{l}{\textit{Executive function} (units = seconds)} \\
\hspace{3mm} Stroop 1 & $\beta_{t10} = 32.2$ (0.19) & $\lambda_{31} = -7.05$  (-) \\
\hspace{3mm} Stroop 2 & $\beta_{t11} = 23.3$ (0.18) & $\lambda_{32} = -3.96$  (0.23) \\
\hspace{3mm} Stroop 3 & $\beta_{t12} = 58.3$ (0.34) & $\lambda_{33} = -20.2$  (0.45) \\
\hspace{3mm} Stroop 4 & $\beta_{t13} = 65.4$ (0.36) & $\lambda_{34} = -21.7$  (0.49) \\
\midrule
\multicolumn{3}{l}{\textit{Retest effects}}  \\
\hspace{3mm} CVLT & $\beta_{r1} = 0.11$ (0.02) & Odds ratio 1.12 \\
\hspace{3mm} Digit span & $\beta_{r2} = 0.07$ (0.02) & Odds ratio 1.08 \\
\hspace{3mm} Stroop conditions 1 and 2 & $\beta_{r3} = -1.24$ (0.23) & - \\
\hspace{3mm} Stroop conditions 3 and 4 & $\beta_{r4} = -2.21$ (0.35) & - \\
\bottomrule
\end{tabular}
\end{threeparttable}
\end{table}

Table \ref{tab:cognition_variance_components} shows the estimated variance components. At level 2, the variance of working memory and executive function were estimated exactly to zero. While it seems implausible that the "true" variances are exactly zero, simulations described in Section \ref{sec:cog_sim} and shown in Figure \ref{fig:cog_sim_smooth_coverage} (right) indicate that zero estimates occur frequently with these data when the ratio of level-2 variance to total level-2 and level-3 variance is low, and we hence take these estimates to indicate that the within-subject variance between timepoints is lower than the between-subject variance. The correlation between levels of the three cognitive abilities varies between 0.32 and 0.43, which is slightly below the meta-analytic results of \citet{tucker-drobCoupledCognitiveChanges2019}, who found a level communality of 0.56 across a large range of cognitive domains.

\begin{table}
\centering
\begin{threeparttable}
\caption{Estimates of variance components in the model presented in Section \ref{sec:cog_app}.}
\label{tab:cognition_variance_components}
\small
\begin{tabular}{lrrr}
\toprule
\multicolumn{4}{l}{\textit{Level 1: Dispersion parameters}}\\
 & CVLT and digit span & Stroop 1+2 & Stroop 3+4  \\
Residual standard error & Fixed to 1 & $\surd\hat{\phi}_{1} = 7.46$ seconds & $\surd\hat{\phi}_{2} = 8.94$ seconds \\
\midrule
\multicolumn{4}{l}{\textit{Level 2: Between-timepoint, within-participant variation}}\\
& Episodic memory & Working memory & Executive function \\
Estimated variance & $\hat{\psi}_{1}^{(2)} = 0.06$ & $\hat{\psi}_{2}^{(2)} = 0$ & $\hat{\psi}_{3}^{(2)} = 0$ \\
\midrule
\multicolumn{4}{l}{\textit{Level 3: Between-participant variation}}\\
 & Episodic memory & Working memory & Executive function \\
Episodic memory & 0.076 & cor = 0.43 & cor = 0.32 \\
Working memory & 0.040 & 0.112 & cor = 0.36 \\
Executive function & 0.043 & 0.059 & 0.237 \\
\midrule
\multicolumn{4}{l}{\textit{Level 4: Spline smoothing}}\\
 & Episodic memory & Working memory & Executive function \\
Estimated variance & $\hat{\psi}_{1}^{(4)} = 5.66 \times 10^{-3}$ & $\hat{\psi}_{2}^{(4)} = 2.04 \times 10^{-3}$ & $\hat{\psi}_{3}^{(4)} = 4.57 \times 10^{-2}$ \\
Smoothing parameter & $\hat{\lambda}_{1} = 1/\hat{\psi}_{1}^{(4)}  = 177$ & $\hat{\lambda}_{2} = 1/\hat{\psi}_{2}^{(4)} = 490$ & $\hat{\lambda}_{3} = 1/\hat{\psi}_{3}^{(4)} = 21.9$ \\
\bottomrule
\end{tabular}
\end{threeparttable}
\end{table}

\begin{figure}
\centering
\includegraphics[width=\columnwidth]{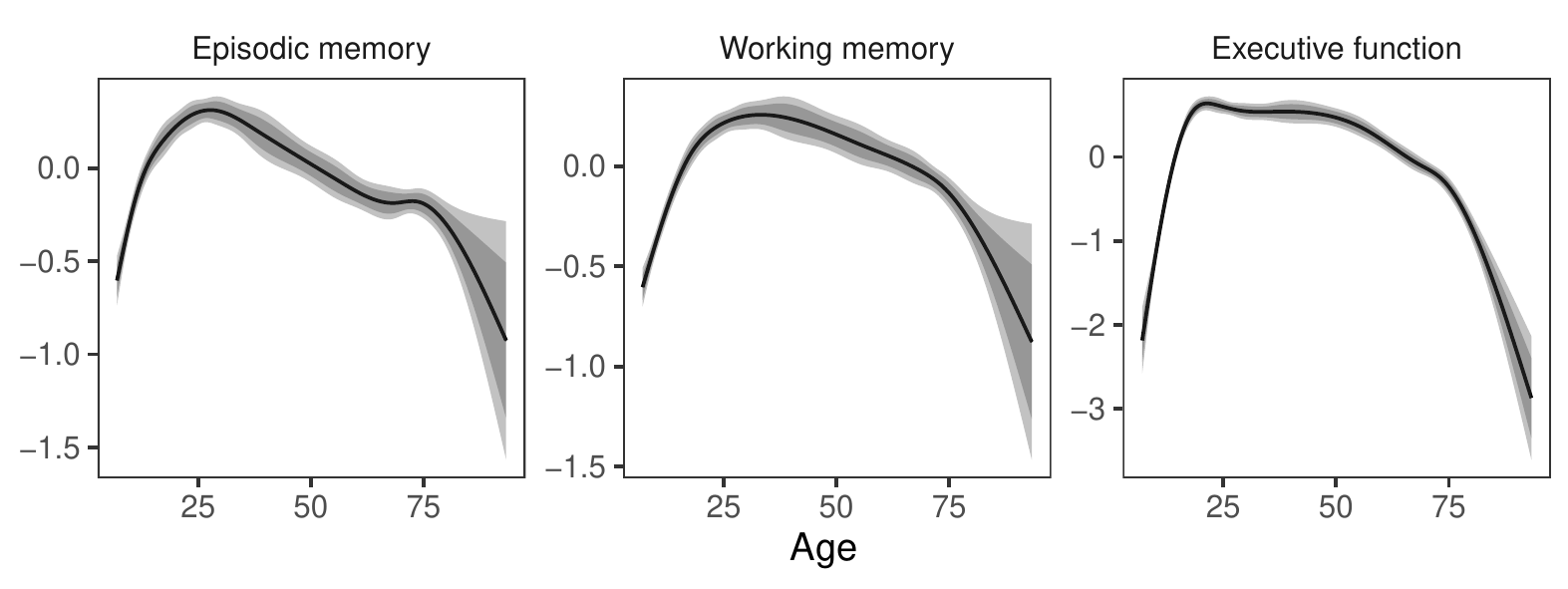}
\caption{\textbf{Estimated lifespan trajectories.} Units on the y-axis are standard deviations of the underlying latent variable $\eta_{m}$. Shaded regions are 95\% pointwise confidence bands (inner) and 95\% simultaneous confidence bands (outer).}
\label{fig:cognition_smooth}
\end{figure}

\begin{figure}
\centering
\includegraphics[width=\columnwidth]{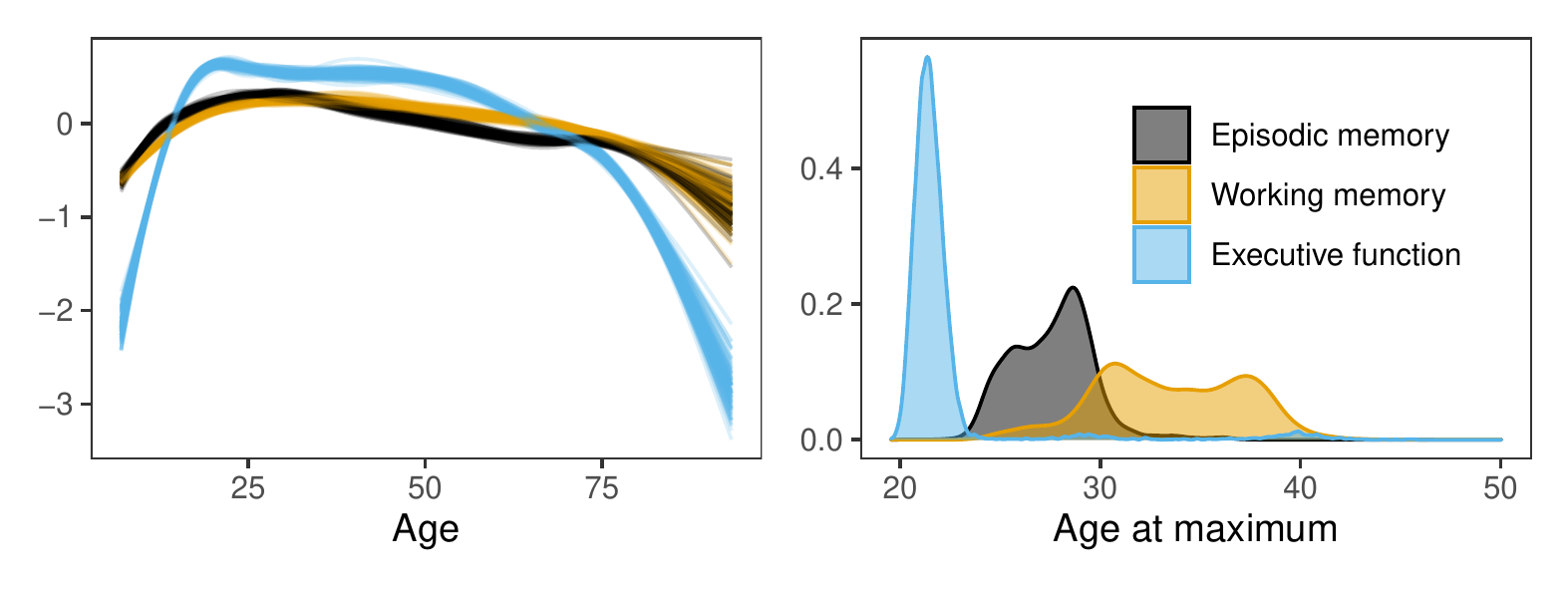}
\caption{\textbf{Empirical Bayes posteriors.} Left: for each cognitive domain, 100 curves for the posterior are shown. Right: posterior densities of the age at which maximum level is attained for each domain.}
\label{fig:cognition_posteriors}
\end{figure}

Figure \ref{fig:cognition_smooth} shows the estimated lifespan trajectories for the three cognitive domains, with pointwise and simultaneous 95\% confidence bands. The latter were obtained by sampling 100,000 spline coefficients from the empirical Bayes posterior distribution, and following the description at the end of Section \ref{sec:GAM_intro} we found critical values $\tilde{z}_{.025}$ close to 3 for all three domains. 100 randomly selected curves for each domain are shown in Figure \ref{fig:cognition_posteriors} (left). The trajectories suggest that executive function reaches its maximum earliest, at the age of 22, while episodic memory peaks at 28 and working memory at 34 years of age. As expected, the curves also indicate a steep increase during childhood, and a steep decrease after about 75 years of age. Given the ceiling effects apparent for CVLT trials in Figure \ref{fig:cognition_data}, some care should be taken should be taken when interpreting the shape of the estimated trajectory for episodic memory, as the test may not be able to discriminate the higher levels of episodic memory. Figure \ref{fig:cognition_posteriors} (right) shows posterior densities for the age associated with maximum ability in each domain. While the posteriors for age at maximum episodic and working memory have some overlap, for executive function the posterior is highly peaked, although with a small additional bump around the age of 40. Table S1 in Online Resource 1 shows the posterior probability of each possible ordering of the age at maximum across the three domains, giving 88.4\% probability to the ordering implied by the point estimate (executive function $<$ episodic memory $<$ working memory), and $88.4 + 6.82 \approx 95.2$\% to the event that executive function has the lowest age at maximum. 

The finding that executive function seems to peak at an earlier age than episodic and working memory is in agreement with previous studies \citep{gajewskiStroopTaskPerformance2020,salthouseExecutiveFunctioningPotential2003,westApplicationPrefrontalCortex1996}. Furthermore, the steady decline after the peak apparent in all three trajectories in Figure \ref{fig:cognition_smooth} is in some agreement with \citet{salthouseWhenDoesAgerelated2009}. On the other hand, the peak in working memory at around 33 years of age does not agree with \citet{gregoireEffectAgeForward1997}, who found the performance on digit span backward and forward tasks to be steadily declining from the age of 16. The peak in episodic memory at around 27 years of age is in some agreement with the cross-sectional results in \citet[Fig. 1]{ronnlundStabilityGrowthDecline2005}, but not with the longitudinal effects adjusted for retest effects from the same study, which suggest a steady level of episodic memory until the age of 60 \citep[Fig. 5]{ronnlundStabilityGrowthDecline2005}. However, all previous studies of the topic which we are aware of have either relied on restricted parametric models, or categorization into discrete age groups, and are hence not directly comparable. The GALAMM-based model presented in this section offers the opportunity for more accurate estimation of lifespan cognitive development, without sacrificing the factor analytic models used to relate multivariate test measurements to a lower number of latent traits.

\subsection{Simulation experiments}
\label{sec:cog_sim}

A parametric bootstrap \citep[Ch. 6.5]{efronIntroductionBootstrap1993} can be used to assess the bias of point estimates and standard errors computed using the proposed maximum marginal likelihood algorithm, by repeatedly sampling new observations from the fitted model. If the marginal log-likelihood \eqref{eq:GLLAMM_laplace_loglik} is regular, i.e., well approximated by a quadratic function in the neighborhood of its maximum, bootstrap standard errors will be close to the standard errors computed from the asymptotic covariance matrix, and accordingly Wald type confidence intervals will have good coverage properties \citep[Ch. 5.2-5.3]{pawitanAllLikelihood2001}. The frequentist interpretation of smoothing via random effects is that each dataset from the population contains a random sample of penalized coefficients, implying that a new curve from the empirical Bayes posterior should be used as the true value in each simulation. If instead viewed as a computational trick to compute maximum marginal likelihood estimates under an empirical Bayes prior, using the point estimate would be appropriate. We here took the latter view. 

When simulating, the data structure, values of all covariates, and parameter estimates were retained, but the linear predictor was updated by sampling new random intercepts $\zeta_{m}^{(2)}$ and $\zeta_{m}^{(3)}$ $(m=1,2,3)$ from normal distributions with covariance components from Table \ref{tab:cognition_variance_components}. New elementary responses were then sampled from the binomial distribution for CVLT and digit span items and from the normal distribution for the Stroop items, and the whole procedure was repeated 500 times.

\begin{figure}
\centering
\includegraphics{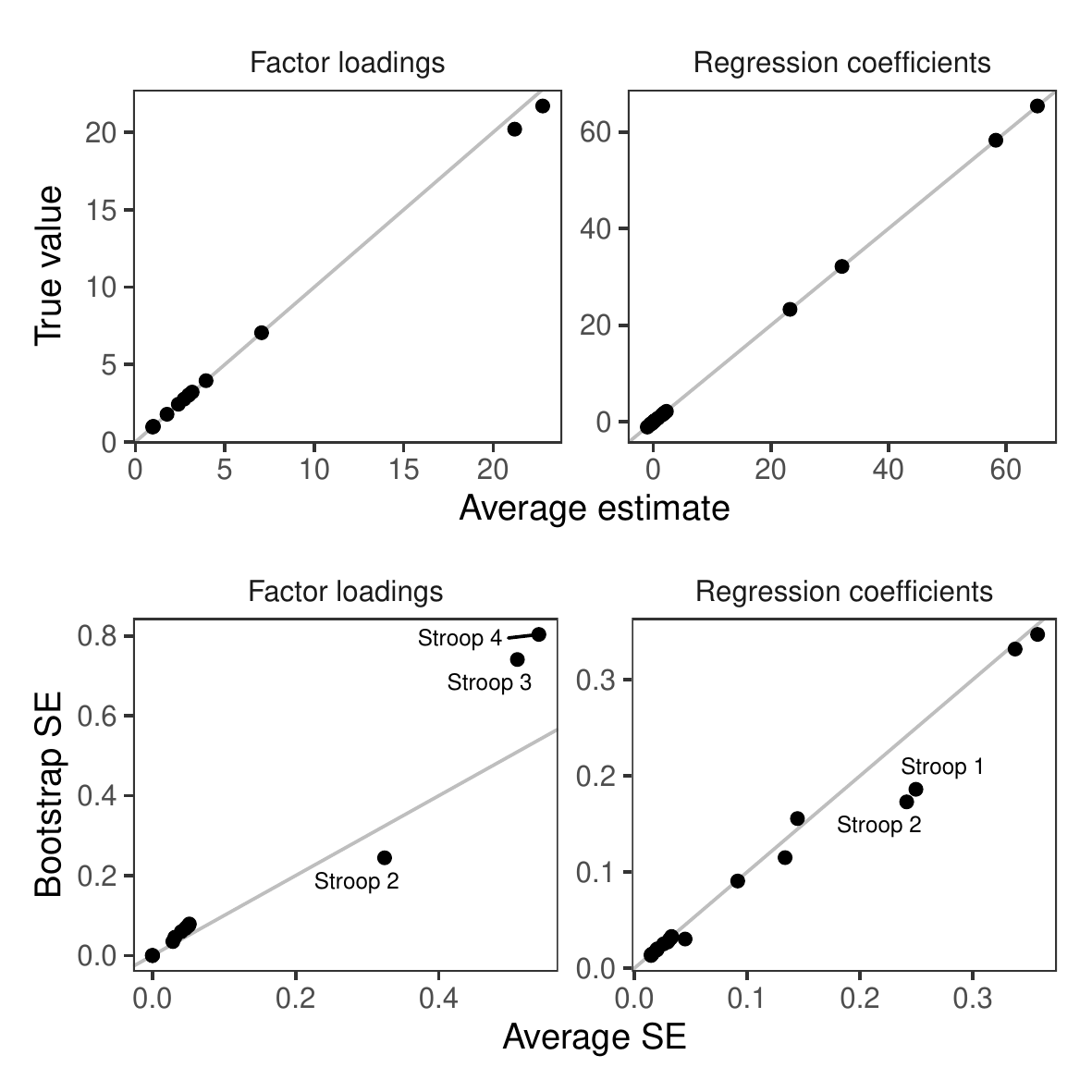}
\caption{\textbf{Bootstrap assessment of bias and standard errors.} The top row shows true values plotted against the simulation averages, and the bottom row shows bootstrap standard error estimates plotted against the average standard errors across bootstrap samples. Outlying observations are labeled.}
\label{fig:cog_sim_param}
\end{figure}

The top row of Figure \ref{fig:cog_sim_param} shows the true values of factor loading and regression coefficients plotted against their average across simulations, indicating that the bias in these terms is close to zero. In the bottom row of Figure \ref{fig:cog_sim_param}, the standard deviation of point estimates of a given parameter across bootstrap samples was plotted against the bootstrap average of the standard error of the same parameter obtained from the asymptotic covariance matrix. For the factor loadings, the bootstrap estimated standard errors for Stroop condition 3 and 4 were larger than the average standard errors, whereas the bootstrap estimate for Stroop condition 2 were lower. Considering the regression coefficients, the bootstrap estimated standard errors for the trial effects of Stroop conditions 1 and 2 were lower than the average standard errors. This means that the profiled marginal log likelihood is not well approximated by a quadratic function for the mentioned parameters, and that confidence intervals for factor loadings should be based on profile likelihood estimation or bootstrapping \citep[Ch. 5.3]{pawitanAllLikelihood2001} (see also \citet[Sec. 3.1.2]{jeonProfileLikelihoodApproachEstimating2012}). This is in agreement with results reported for other mixed models, e.g., \citet{boothBootstrapMethodsGeneralized1995}, \citet{brockwellComparisonStatisticalMethods2001}, and \citet[Sec. 3.4]{demidenkoMixedModelsTheory2013}. For all other parameters, the bottom row of Figure \ref{fig:cog_sim_param} shows that the average standard errors were close to their bootstrap counterparts. Bootstrap and asymptotic confidence intervals for all regression coefficients and factor loadings are reported in Tables S2 and S3 of Online Resource 1.

Figure S2 in Online Resource 1 shows that the average estimates of the smooth functions were almost overlapping with the true functions. In units of standard deviations of the latent variable $\eta_{m}$, the root-mean-square error over bootstrap estimates was 0.041 for episodic memory, 0.058 for working memory, and 0.081 for executive function. For comparison, the range (difference between maximum and minimum) of the trajectories over the lifespan were 1.73, 1.96, and 4.94 standard deviations, respectively. The three trajectories shown in Figure \ref{fig:cognition_smooth}, which were the ground truth in the simulation experiments, had total effective degrees of freedom equal to 24.6. In contrast, the average effective degrees of freedom over the bootstrap samples was 20.4, and only on two occasions did it exceed 24.6\footnote{A histogram of effective degrees of freedom across bootstrap samples is shown in Figure S3 of Online Resource 1.}. This confirms that the maximum marginal likelihood estimation protects against overfitting by yielding estimates which (for finite samples) are smoother than the data generating function, as expected by the results of \citet{reissSmoothingParameterSelection2009} and \citet{woodFastStableRestricted2011}.

\begin{figure}
\centering
\includegraphics[width=.49\columnwidth]{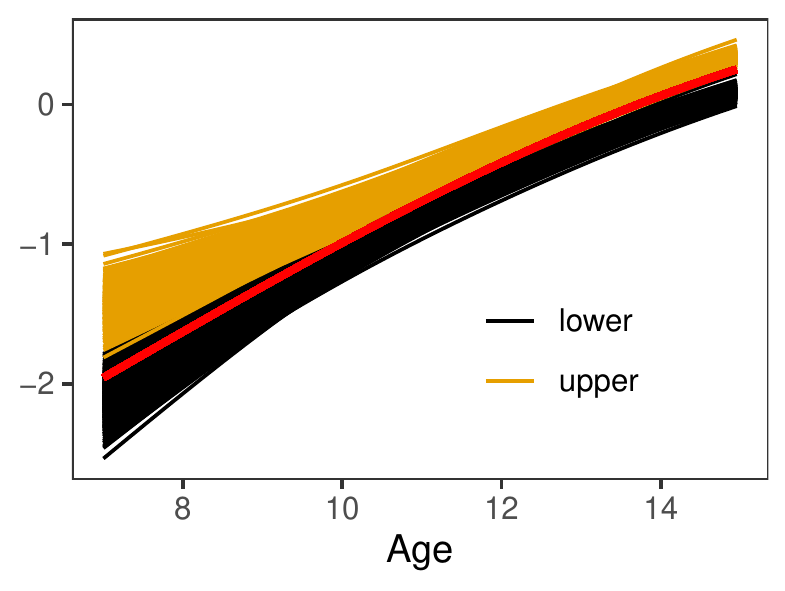}
\includegraphics[width=.49\columnwidth]{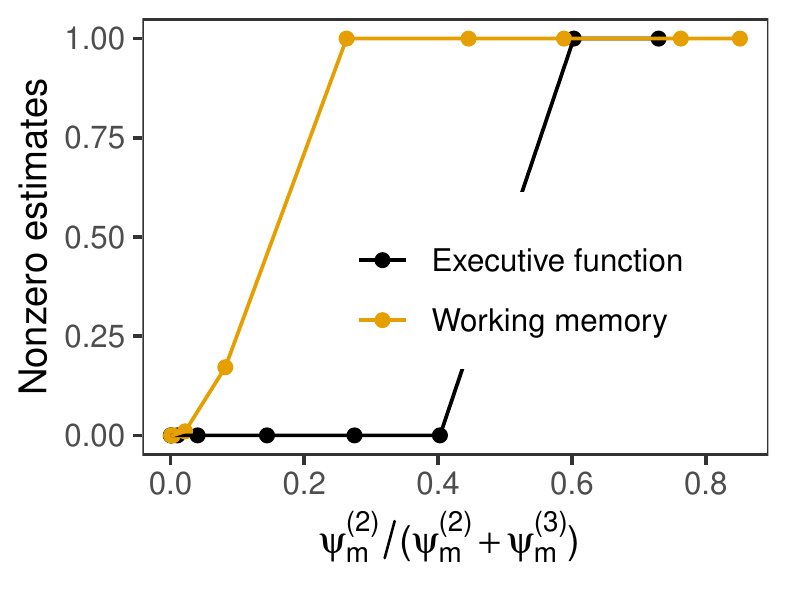}
\caption{\textbf{Simultaneous results.} Left: All bootstrap samples of lower and upper simultaneous confidence bands plotted together with the true function in read, for ages between 7 and 15 years. Right: Proportion of $\hat{\psi}_{m}^{(2)}$ obtaining a nonzero estimate as the true value increases, for working memory ($m=2$) and executive function ($m=3$). The x-axis shows the ratio of level-2 variance to total level-2 and level-3 variance.}
\label{fig:cog_sim_smooth_coverage}
\end{figure}

The across-the-function coverage of pointwise 95\% confidence bands were conservative for episodic and working memory with 100\% and 98\% coverage, but too liberal for executive function, with an average of 91\% coverage. The simultaneous 95\% confidence bands contained the true function with almost 100\% probability for episodic memory and 98\% probability for working memory, but only 82\% probability for executive function. Figure \ref{fig:cog_sim_smooth_coverage} (left) shows one source of the poor simultaneous coverage for executive function: the true function (in red) is below a sizeable proportion of the lower simultaneous confidence bands for ages below 10. Also here bootstrapping would likely yield better coverage properties \citep{hardleBootstrappingNonparametricRegression1988,hardleBootstrapSimultaneousError1991,hardleBOOTSTRAPINFERENCESEMIPARAMETRIC2004}, and in this case 95\% bootstrap confidence bands did contain the true function over the full range. However, addressing the coverage of bootstrap based confidence bands over a range of datasets sampled from the population is beyond the scope of this paper.

We finally investigated the level-2 variances variance for working memory and executive function, which were estimated to be exactly zero in the previous section, cf. Table \ref{tab:cognition_variance_components}. We gradually increased the value of $\psi_{m}^{(2)}$, $m=2,3$, otherwise simulating data as before, and recorded the proportion of simulated samples for which these variance parameters were estimated to zero. The results are shown Figure \ref{fig:cog_sim_smooth_coverage} (right). For working memory, the level-2 variance was given a nonzero estimate in all simulated samples already when the level-2 variance reach 20\% of the total variance. In contrast, the level-2 variance for executive function was estimated to zero until it reached half the total variance.

\section{Latent covariates}
\label{sec:latent_covariates}

\subsection{Socioeconomic status and hippocampus volume}
\label{sec:ses_model}

\begin{figure}
\centering
\includegraphics[width=.49\columnwidth]{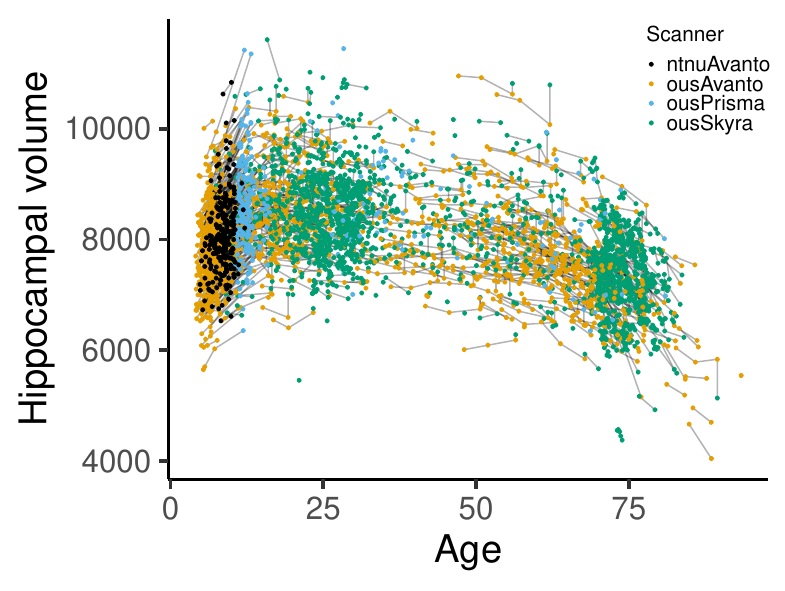}
\includegraphics[width=.49\columnwidth]{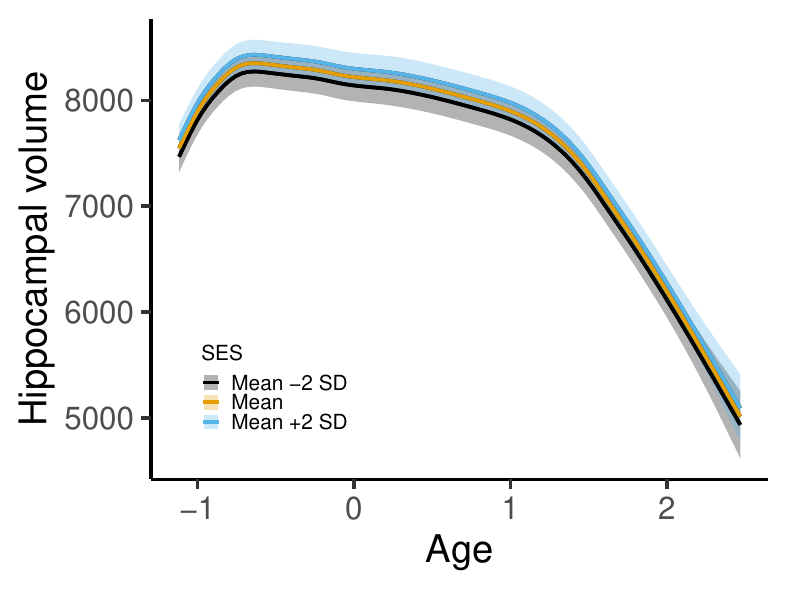}
\caption{\textbf{Hippocampal volume curves.} Left: Total volumes of left and right hippocampus (in mm$^{3}$) plotted versus age. Repeated observations of the same individual are connected with gray lines. Right: Estimated hippocampal volume trajectories at mean socioeconomic status (SES) and at two standard deviation above or below mean. Shaded regions show 95\% pointwise confidence intervals for SES two standard deviations above or below mean.}
\label{fig:hippocampus}
\end{figure}

The association between socioeconomic status and brain development has been the subject of much research. It has been proposed that higher socioeconomic status protects against late-life dementia \citep{livingstonDementiaPreventionIntervention2017}, whereas a meta-analysis found that the associations between socioeconomic status and brain structure varied considerably between samples \citep{walhovdEducationIncomeShow2021}. The hippocampus is a brain region which plays an important role in memory consolidation, and is one of the first regions to be damaged in Alzheimer's disease \citep{duboisPreclinicalAlzheimerDisease2016}. Positive associations have been found between socioeconomic status and hippocampal volume in children \citep{hansonAssociationIncomeHippocampus2011,nobleNeuralCorrelatesSocioeconomic2012,nobleFamilyIncomeParental2015,yuSocioeconomicStatusHippocampal2018}, and between childhood socioeconomic status and adult brain size \citep{staffChildhoodSocioeconomicStatus2012}. However, while hippocampal volume is known to be a nonlinear function of age, most studies investigating the association have used linear regression analyses. An exception is \citet{nybergEducationalAttainmentDoes2021}, who used GAMMs to model the hippocampal trajectory, and found evidence for a close-to-zero association between longitudinal change in hippocampal volume and educational attainment in two large adult samples.

We here consider the association between hippocampal volume and socioeconomic status across the lifespan, still using data from the Center for Lifespan Changes in Brain and Cognition \citep{fjellNeuroinflammationTauInteract2018,walhovdNeurodevelopmentalOriginsLifespan2016}. Hippocampal volumes were estimated with FreeSurfer 7 \citep{daleCorticalSurfaceBasedAnalysis1999,fischlWholeBrainSegmentation2002,reuterWithinsubjectTemplateEstimation2012} from magnetic resonance images obtained at four different scanners, and are shown in Figure \ref{fig:hippocampus} (left). In total, we had 4248 scans of 1916 participants aged between 4 and 93 years, with between 1 and 8 scans per participant. Our interest concerns how the lifespan trajectory of hippocampal volume depends on socioeconomic status. For participants below the age of twenty, we defined socioeconomic status based on their father's and mother's years of completed education and income, and for participants above the age of twenty we defined it based on their own education and income. As these variables were typically only measured at a single timepoint, they were considered time-independent. Of the 1916 participants with hippocampal volume measurements, either their own or at least one parent's education level was available from 1661 participants, while the corresponding number for income was 571\cprotect\footnote{One might debate whether the measured items \emph{reflect} socioeconomic status or whether socioeconomic status instead is \emph{formed} by the measured items, see \citet[p. 67]{skrondalGeneralizedLatentVariable2004} and \citet{edwardsNatureDirectionRelationships2000}. In this example we assume a \emph{reflective} model.}. All timepoints for 253 participants with no measurement of socioeconomic status were also included in the analyses, yielding a total of 7264 level-1 units.

Since all outcomes were continuous, we used a unit link function and measurement model
\begin{equation}
\label{eq:ses_measurement}
y_{i} = \mathbf{d}_{\text{s},i}'\boldsymbol{\beta}_{\text{s}} + d_{h,i}\left( \mathbf{x}_{\text{h},i}' \boldsymbol{\beta}_{\text{h}} + f\left(a_{i}\right)\right) +  \eta_{1} \mathbf{z}_{i}' \boldsymbol{\lambda} + d_{\text{h},i} \eta_{2} + \epsilon_{i},
\end{equation}
where $\boldsymbol{\beta}_{\text{s}}$ contains the intercepts for the items measuring socioeconomic status and $\mathbf{d}_{\text{s},i}$ is a vector of length 6 whose $k$th element is an indicator for the event that the $i$th level-1 unit measures the $k$th socioeconomic status item. Variable $d_{h,i} \in \{0,1\}$ indicates whether the $i$th level-1 units is a measurement of hippocampal volume, $\mathbf{x}_{\text{h},i}$ is a vector of linear regression terms for scanner, sex, and intracranial volume, and $\boldsymbol{\beta}_{\text{h}}$ are corresponding regression coefficients\cprotect\footnote{\citet{alfaro-almagroConfoundModellingUK2021} and \citet{hyattQuandaryCovaryingBrief2020} give overviews of variables to control for in analysis of neuroimaging data.}. The age of the participant to which the $i$th level-1 unit belongs is denoted $a_{i}$, and $f(\cdot)$ is a smooth function composed as a linear combination of fifteen cubic regression splines, subject to sum-to-zero constraints as described in \citet[Ch. 5.4.1]{woodGeneralizedAdditiveModels2017a}. Latent socioeconomic status is represented by $\eta_{1}$, and $\boldsymbol{\lambda} = (\lambda_{1}, \dots ,\lambda_{8})^{T}$ is a vector of factor loadings. Factor loadings for paternal, maternal, and the participant's own education level were represented by $\lambda_{1}, \dots, \lambda_{3}$, and the corresponding factor loadings for paternal, maternal, and the participant's own income were represented by $\lambda_{4},\dots,\lambda_{6}$. Accordingly, when the $i$th level-1 unit is a measurement of income or education, $\mathbf{z}_{i}$ is an indicator vector which ensures that the correct factor loading among $\lambda_{1},\dots, \lambda_{6}$ is multiplied by $\eta_{1}$. Finally, $\lambda_{7}$ represented the effect of latent socioeconomic status on hippocampal volume, and $\lambda_{8}$ the interaction effect of age and socioeconomic status on hippocampal volume. Hence, when the $i$th level-1 unit is a measurement of hippocampal volume, $\mathbf{z}_{i}^{T} = (0,\dots,0, 1, a_{i})$. Since the data contained repeated scans, a random intercept for hippocampal volume $\eta_{2}$ was also included. A heteroscedastic model for the residuals was assumed, $\epsilon_{i} \sim N(0, \sigma_{g(i)}^{2})$, where $g(i)=1$ if the $i$th level-1 unit is a measurement of income, $g(i)=2$ if it is a measurement of education level, and $g(i)=3$ if it is a measurement of hippocampal volume. The structural model was simply $\boldsymbol{\eta} = \boldsymbol{\zeta} \sim N(\mathbf{0}, \boldsymbol{\Psi})$ where $\boldsymbol{\Psi} = \text{diag}(\psi_{1}, \psi_{2})$. Assuming zero correlation between level-2 disturbances was required for identifiability, since $\eta_{2}$ depends on $\eta_{1}$ through $\lambda_{7}$ and $\lambda_{8}$. The proportion of structural zeroes in the random effects design matrix $\mathbf{Z}$ was 99.8 \%.

Since we used a unit link function and normally distributed residuals, the Laplace approximation was exact. Income and education variables were log-transformed to obtain response values closer to a normal distribution. When fitting the models, all quantitative variables were transformed to have zero mean and unit standard deviations. For identifiability, $\lambda_{1}$ was fixed to unity on the transformed scale used in model fitting.

The model described above has seven free factor loadings, $\lambda_{2},\dots,\lambda_{8}$, and we compared it to constrained versions using the marginal Akaike information criterion (AIC) defining the model degrees of freedom by the number of parameters \citep{akaikeNewLookStatistical1974,vaidaConditionalAkaikeInformation2005}. Based on the results shown in Table \ref{tab:ses_aic} we chose model (f), with equal loadings for the education items, equal loadings for the income items, and no interaction effect between age and socioeconomic status on hippocampal volume.

\begin{table}
\centering
\begin{threeparttable}
\caption{Comparison of models for the effect of socioeconomic status on hippocampal volume. AIC values have been shifted to be zero for the full model, for ease of comparison. }
\label{tab:ses_aic}
\small
\begin{tabular}{llll}
\toprule
Model & Parameters & Log-likelihood & AIC \\
\midrule
(a): Free loadings & 26 & -5574 & 0.00 \\
(b): (a) and no interaction, $\lambda_{8}=0$ & 25 & -5575 & -0.38 \\
(c): Parents equal, $\lambda_{1}=\lambda_{2}$ and $\lambda_{4}=\lambda_{5}$ & 24 & -5576 & -1.29 \\
(d): (c) and no interaction, $\lambda_{8}=0$ & 23 & -5577 & -1.64 \\
(e): Item groups equal, $\lambda_{1}=\lambda_{2}=\lambda_{3}$ and $\lambda_{4}=\lambda_{5}=\lambda_{6}$ & 22 & -5576 & -5.22 \\
(f): (e) and no interaction, $\lambda_{8}=0$ & 21 & -5577 & -5.58 \\
(g): (f) and no main effect, $\lambda_{7}=\lambda_{8}=0$ & 20 & -5578 & -4.18 \\
\bottomrule
\end{tabular}
\end{threeparttable}
\end{table}

\begin{table}
\centering
\begin{threeparttable}
\caption{Parametric terms in model of hippocampal volume and socioeconomic status.}
\label{tab:ses_parametric_terms}
\small
\begin{tabular}{lllll}
\toprule
Parameter & Estimate & SE & Units \\
\midrule
\multicolumn{4}{l}{\textit{Effects on hippocampal volume}}\\
\hspace{3mm} Scanner ousAvanto, $\beta_{h1}$ & -72.2 & 57.5 & mm$^{3}$ \\
\hspace{3mm} Scanner ousPrisma, $\beta_{h2}$ & 80.9 & 64.5 & mm$^{3}$ \\
\hspace{3mm} Scanner ousSkyra, $\beta_{h3}$ & 248 & 58.5 & mm$^{3}$ \\
\hspace{3mm} Total intracranial volume, $\beta_{h4}$ & 0.00201 & $9.05 \times 10^{-5}$ & mm$^{3}$/mm$^{3}$ \\
\hspace{3mm} Sex=Male, $\beta_{h5}$ & 217 & 32.9 & mm$^{3}$ \\
\midrule
\multicolumn{4}{l}{\textit{Factor loadings}}\\
\hspace{3mm} Education, $\lambda_{1}=\lambda_{2}=\lambda_{3}$ & 0.168 & - &  $\log(\text{years})$ \\
\hspace{3mm} Income, $\lambda_{4}=\lambda_{5}=\lambda_{6}$ & 0.266 & 0.0448 &  $\log(\text{NOK})$ \\
\hspace{3mm} Hippocampus, $\lambda_{7}$ & 59.1 & 32 &  mm$^{3}$ \\
\midrule
\multicolumn{4}{l}{\textit{Variance components}}\\
\hspace{3mm} Socioeconomic status, $\surd\psi_{1}$ & 0.669 & - &  - \\
\hspace{3mm} Hippocampus, $\surd\psi_{2}$ & 601 & - &  mm$^{3}$ \\
\hspace{3mm} Income residual, $\sigma_{1}$ & 0.593 & - &  $\log(\text{NOK})$ \\
\hspace{3mm} Education residual, $\sigma_{2}$ & 0.125 & - &  $\log(\text{years})$ \\
\hspace{3mm} Hippocampus residual, $\sigma_{3}$ & 134 & - &  mm$^{3}$ \\
\bottomrule
\end{tabular}
    \begin{tablenotes}
      \footnotesize
      \item{}NOK denotes Norwegian kroner, with 10 NOK$\approx$1 EUR. Scanner effects are relative to 'ntnuSkyra', see Figure \ref{fig:hippocampus} (left). Units mm$^{3}$/mm$^{3}$ for total intracranial volume represent mm$^{3}$ of hippocampus per mm$^{3}$ of total intracranial volume. The factor loading for education does not have a standard error, as it was fixed for identifiability.
    \end{tablenotes}
\end{threeparttable}
\end{table}

Table \ref{tab:ses_parametric_terms} shows the estimated parametric effects of main interest. Item intercepts $\boldsymbol{\beta}_{s}$ can be found in Table S1 of Online Resource 2. Standard errors are not reported for variance components, as their likelihood is typically not regular. As expected, higher total intracranial volume and being male were associated with higher hippocampal volume \citep{hyattQuandaryCovaryingBrief2020}. From the estimated standard deviation of the random intercept for hippocampal volume and the residual standard deviation for hippocampal volume, we find an intraclass correlation (ICC) of $=601^{2}/(601^{2}+134^{2}) = 0.95$. An ICC this high implies that the variation between individuals is much larger than the variation between different timepoints of the same individual, as is also clear from the raw data plot in Figure \ref{fig:hippocampus} (left). The estimated factor loading for income was positive, with two-sided $p$-value $2 \times 10^{-8}$ computed using a likelihood ratio test as described in the next paragraph, indicating that both education and income are positively related to the latent construct $\eta_{1}$. It also follows that the difference between mean socioeconomic status and a socioeconomic status one standard deviation above the mean  is associated with a difference in education level of $\exp(\hat{\beta}_{s3} + \hat{\lambda}_{3} \surd\hat{\psi}_{1}) - \exp(\hat{\beta}_{s3}) = 2$ years and with difference in annual income of $\exp(\hat{\beta}_{s6} + \hat{\lambda}_{6} \surd\hat{\psi}_{1}) - \exp(\hat{\beta}_{s6}) = 95 \times 10^{3}$ NOK, where we have taken $\hat{\beta}_{s3}=2.81$ and $\hat{\beta}_{s6}=13.1$ from Table S1 of Online Resource 2. Note that this effect is not additive on the natural scale, since education and income levels were log-transformed.

Figure \ref{fig:hippocampus} (right) shows the estimated hippocampal trajectories at three levels of socioeconomic status. We tested the null hypothesis of no effect of socioeconomic status on hippocampal volume using a likelihood ratio test. In particular, under the null hypothesis, twice the difference between the log-likelihoods of model (f) and model (g) in Table \ref{tab:ses_aic} is distributed according to a $\chi^2$-distribution with one degree of freedom (e.g., \citet[Sec. 8.3.4]{skrondalGeneralizedLatentVariable2004}). The resulting $p$-value was $0.065$, thus not significant at a 5\% level. From the point estimate, we see that a one standard deviation increase in socioeconomic status is associated with a $\hat{\lambda}_{7}\surd\hat{\psi}_{1} = 40 \text{ mm}^{3}$ increase in hippocampal volume. For comparison, the rate of increase seen during childhood in Figure \ref{fig:hippocampus} (left) is around 50 mm$^{3}$/year, the rate of decline during adulthood around 10-15 mm$^{3}$/year, increasing to 90-100 mm$^{3}$/year in old age. Assuming no birth cohort effects \citep{baltesLongitudinalCrossSectionalSequences1968} and representative sampling, the presence of a constant effect $\lambda_{7}$ and the absence of an interaction effect $\lambda_{8}$, would imply that socioeconomic status affects early life brain development, rather than the rate of change at any point later in life. However, this analysis is inconclusive with regards to such a hypothesis.

\subsection{Simulation experiments}
\label{sec:ses_simulation}

Simulation experiments were performed based on the model estimated in the previous section. In particular, we were interested in understanding model selection with AIC as performed in Table \ref{tab:ses_aic} and the estimation of hippocampal volume trajectories as in Figure \ref{fig:hippocampus} (right). To this end, we simulated data using estimated model parameters and a data structure closely following the real data, as shown in Online Resource 2, Figure S1. For simplicity, explanatory variables related to scanner, sex, and intracranial volume were not included in the simulations, but otherwise the model was identical to \eqref{eq:ses_measurement}, with parameter values reported in Table \ref{tab:ses_parametric_terms} and Table S1 of Online Resource 2. The simulations were repeated with six discrete values of the interaction parameter $\lambda_{8}$, ranging from 0 to 0.12. Zero interaction implies that the trajectories for different levels of socioeconomic status are parallel, as in Figure \ref{fig:hippocampus} (right), whereas a positive interaction implies that high socioeconomic status is associated with a lower rate of aging in adulthood. This is illustrated in Figure S2 of Online Resource 2. For all parameter settings, 500 Monte Carlo samples with 1916 participants were randomly sampled, and models corresponding to (e) and (f) in Table \ref{tab:ses_aic} were fitted.

\begin{figure}
\centering
\includegraphics[width=.49\columnwidth]{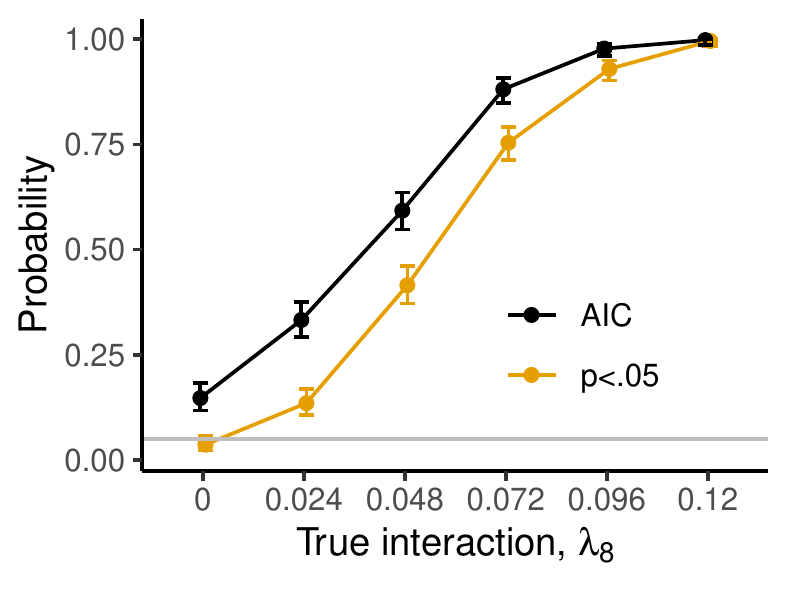}
\includegraphics[width=.49\columnwidth]{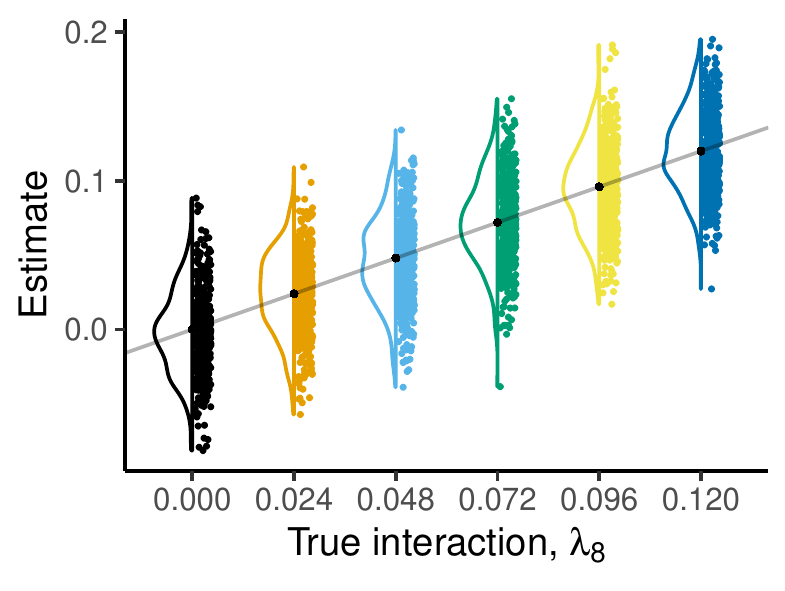}
\caption{\textbf{Interaction term in latent covariates model.} Left: Probability of selecting a model containing an interaction term as a function of the magnitude of the interaction. 'AIC' denotes Akaike information criterion and '$p<.05$' denotes selection based on testing $\lambda_{8}=0$ versus $\lambda_{8}>0$. Error bars show 95\% confidence intervals. The horizontal gray lines shows the $p=0.05$ level, for reference. Right: Violin-dotplots \citep{hintzeViolinPlotsBox1998} of estimated interactions for different values of the true interaction. Gray line and black points indicate the true values, and colored points indicate estimates in single Monte Carlo samples. Values are based on 500 Monte Carlo samples for each parameter combination.}
\label{fig:ses_sim_parametric}
\end{figure}

Figure \ref{fig:ses_sim_parametric} (left) shows results of comparing models (e) and (f) in Table \ref{tab:ses_aic} with AIC and a likelihood ratio test. With true interaction zero, the probability of falsely rejecting the null hypothesis $\lambda_{8}=0$ was close to nominal, and the probability of AIC selecting the model containing this interaction term was close to the expected value of 16\%. Furthermore, the curves suggest that we would have around 80\% power to detect a moderate interaction $\lambda_{8}\approx 0.08$. Figure \ref{fig:ses_sim_parametric} (right) shows the distribution of estimates $\hat{\lambda}_{8}$ in the larger model (e) over all Monte Carlo samples. It is evident that the estimated interactions are symmetrically distributed around their true values. The estimates had low bias also for the other factor loadings, except for the estimates of $\lambda_{7}$ under the misspecified model (f) when the true $\lambda_{8}$ was nonzero, cf. Figure S3 of Online Resource 2. 

\begin{figure}
\centering
\includegraphics[width=\columnwidth]{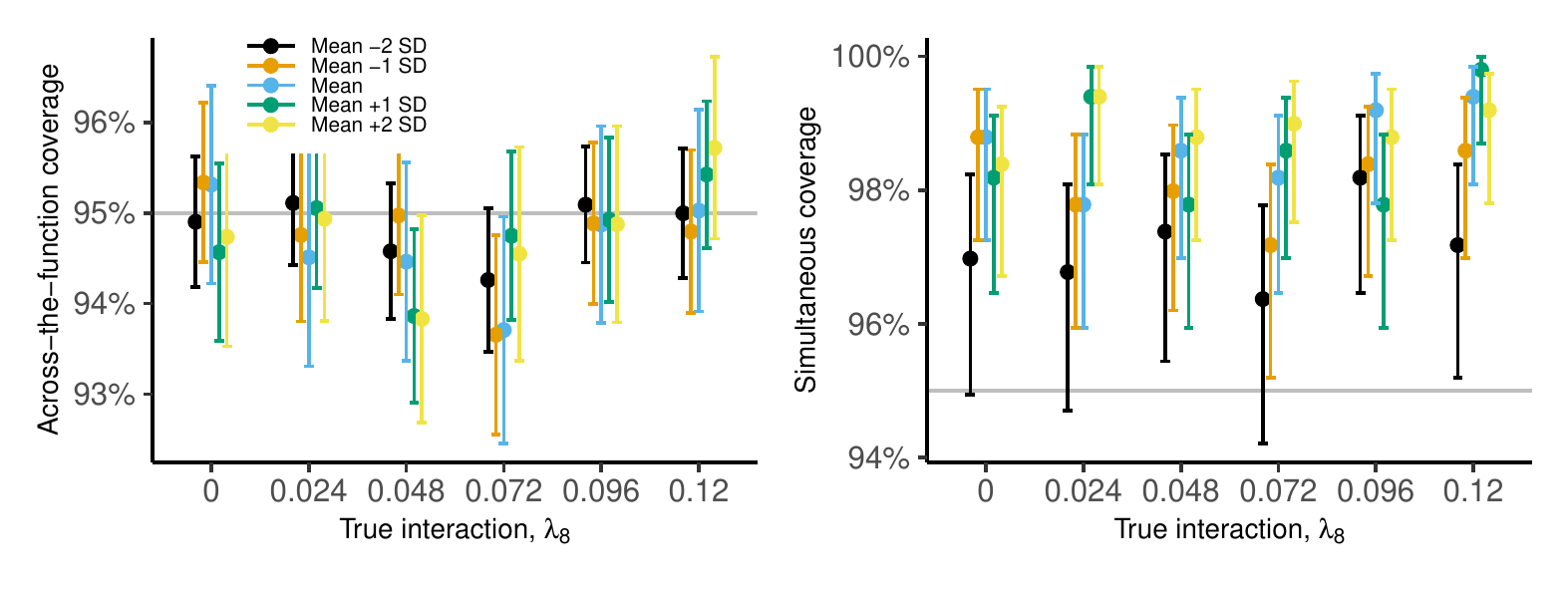}
\caption{\textbf{Coverage of smooth terms in latent covariates model.} Across-the-function coverage of pointwise confidence intervals (left) and coverage of simultaneous confidence intervals (right) for five levels of latent socioeconomic status $\boldsymbol{\eta}_{1}$. Intervals were computed with model (e), which contained a non-zero interaction term $\lambda_{8}$. Error bars show 95\% confidence intervals for simulation estimates.}
\label{fig:ses_sim_smooth}
\end{figure}

Finally, we investigated confidence bands for lifespan trajectories at latent socioeconomic status equal to the mean or one or two standard deviations above or below mean, corresponding to the curves in Figure \ref{fig:hippocampus} (right). As shown in Figure \ref{fig:ses_sim_smooth}, pointwise confidence bands had close to nominal coverage, whereas simultaneous confidence bands in general were conservative, with coverage above 95\%.

\section{Discussion}
\label{sec:discussion}

We have proposed the GALAMM framework for multilevel latent variable modeling, which combines SEM and item response models' ability to model a measurement process with GAMs' ability to flexibly estimate smooth functional relationships. By transforming the GALAMM to mixed model form, the smoothing parameters become inverse variance components which can be estimated jointly with all other model parameters, using maximum marginal likelihood. Possible applications beyond those presented in this paper include spatial smoothing for analysis of regional variations in attitudes measured by social surveys \citep{fahrmeirBayesianSemiparametricLatent2007}.

The latent response model used in Section \ref{sec:latent_response} accommodates a mix of continuous and discrete responses, inducing dependence between latent responses of interest through the latent variable distributions. This approach was inspired by GLMMs \citep{faesHighdimensionalJointModel2008,fieuwsPairwiseFittingMixed2006,iddiJointMarginalizedMultilevel2012,ivanovaMixedModelsApproaches2016} and GLLAMMs \citep[Ch. 14]{skrondalGeneralizedLatentVariable2004} for mixed response types discussed in the literature previously. Several extensions of the model are possible. The assumption of age-invariant measurements could be relaxed with age-dependent factor loadings, yielding a non-uniform differential item functioning model \citep{swaminathanDetectingDifferentialItem1990}. With a higher number of timepoints per individual, inclusion of random slopes would allow estimation of how individual change is correlated across cognitive domains as well as level-slope correlation within domains. These topics were studied in a recent meta analysis \citep{tucker-drobCoupledCognitiveChanges2019} in which all the contributing studies had analyzed samples of adults using linear models. GALAMM would more easily allow such studies of coupled cognitive change across the lifespan, since the nonlinear effect of age is flexibly handled by smooth terms. The simulation studies in Section \ref{sec:cog_sim} suggest that regularity of the likelihood function should be carefully checked before computing Wald type confidence intervals. The bootstrap procedure provides a natural way of checking this, albeit at a high computational cost. The simulations also revealed some weak points worthy of further investigation. Firstly, simultaneous confidence bands for the lifespan trajectory of executive function had too low coverage. A potential way of improving this is by incorporating smoothing parameter uncertainty into the empirical Bayes posterior distribution used to compute the simultaneous intervals, as has been demonstrated by \citet{woodSmoothingParameterModel2016} for GAMs. Alternatively, simultaneous confidence bands can be computed using the bootstrap as demonstrated in Section \ref{sec:cog_sim} \citep{hardleBootstrappingNonparametricRegression1988,hardleBootstrapSimultaneousError1991,hardleBOOTSTRAPINFERENCESEMIPARAMETRIC2004}, albeit at a much increased computational cost. Secondly, the level-2 (within-subject between-timepoint) variances of working memory and executive function were estimated exactly to zero, and as shown by the simulation experiments reported in Figure \ref{fig:cog_sim_smooth_coverage} (right), this will happen for the given data structure when the level-2 variances are relatively small compared to the total level-2 and level-3 variance. This inaccuracy might be due to the Laplace approximation used for computing the marginal likelihood, which has been shown to work poorly for certain models with binomial responses \citep{joeAccuracyLaplaceApproximation2008}. More accurate integral approximations can be obtained with adaptive Gauss-Hermite quadrature \citep{cagnoneLatentVariableModels2013,pinheiroApproximationsLogLikelihoodFunction1995,pinheiroEfficientLaplacianAdaptive2006,rabe-heskethReliableEstimationGeneralized2002,rabe-heskethMaximumLikelihoodEstimation2005}, which unfortunately is not directly suited for data with crossed random effects, although \citet{ogdenSequentialReductionMethod2015}'s sequential reduction method might alleviate this. Alternatively, the Laplace approximation can be improved by retaining more terms in the Taylor expansion \eqref{eq:GLLAMM_integrand_taylor} \citep{anderssonEstimationLatentRegression2021,demidenkoMixedModelsTheory2013,raudenbushMaximumLikelihoodGeneralized2000}. Both these methods for improving the approximation of the integral \eqref{eq:GLLAMM_marginal_likelihood} have a higher computational cost than the Laplace approximation, and developing scalable and more accurate algorithms remains an important topic for further research.

The latent covariates model in Section \ref{sec:latent_covariates} could be further extended be investigating the effect of socioeconomic status on a larger set of brain regions. If supported by domain knowledge, increased power in such a model could be obtained with a factor-by-curve interaction model \citep{coullSimpleIncorporationInteractions2001}, in which the trajectories are assumed to have similar shape and/or smoothness across brain regions. An excellent overview of such hierarchical GAMs is given in \citet{pedersenHierarchicalGeneralizedAdditive2019}. Factor analytic models have also been used for integrating multiple measurements of brain structural integrity \citep{dahlIntegrityDopaminergicNoradrenergic2022,kohnckeHippocampalParahippocampalGray2020} or volumes in the left and right hemispheres \citep{dahlRostralLocusCoeruleus2019}, all of which can be directly incorporated in the proposed framework. In Section \ref{sec:latent_covariates} we used marginal AIC for selecting parametric fixed effects. For selecting smooth terms, on the other hand, conditional AIC with correction for smoothing parameter uncertainty would be appropriate \citep{grevenBehaviourMarginalConditional2010,saefkenUnifyingApproachEstimation2014,woodSmoothingParameterModel2016,yuConditionalAkaikeInformation2012}. For GAMs, \citet[Sec. 4]{woodSmoothingParameterModel2016} show how the covariance matrix of the log smoothing parameter can be used to define a corrected conditional AIC for this purpose, but for use with GALAMMs this approach would need to be implemented with sparse matrix methods.

An interesting extension of the framework is to allow smooth functions to depend on latent variables. This leads to a product of normally distributed latent variables in the mixed model representation, and computing the marginal likelihood \eqref{eq:GLLAMM_marginal_likelihood} thus involves integrating over variables distributed according to the generalized chi-squared distribution, making the Laplace approximation \eqref{eq:GLLAMM_laplace_loglik} inappropriate. The algorithm proposed by \citet{rockwoodMaximumLikelihoodEstimation2020} provides an efficient solution for the case of two-level SEMs with random slopes of latent covariates and normally distributed responses, by first reducing the dimension of the integral and then using Gaussian quadrature for integral approximation. A different approach to a related problem is given by \citet{ganguliAdditiveModelsPredictors2005}, who considered single-level semiparametric models with measurement error in the smooth terms, and used an EM algorithm to correct for measurement error bias. The stochastic approximation EM algorithm \citep{delyonConvergenceStochasticApproximation1999} has also been succesfully applied to estimation of nonlinear mixed models involving intractable integrals \citep{cometsParameterEstimationNonlinear2017,kuhnMaximumLikelihoodEstimation2005}, and may be possible to extend to the models considered in this paper.

The algorithm for maximum marginal likelihood estimation presented in Section \ref{sec:Estimation} was mainly inspired by the sparse matrix methods developed for linear mixed models by \citet{batesFittingLinearMixedEffects2015} and the algorithm proposed by \citet{pinheiroEfficientLaplacianAdaptive2006} for estimating GLMMs with nested random effects. The main extension in our approach involves mapping the factor loadings $\boldsymbol{\lambda}$ and regression coefficients $\mathbf{B}$ to the matrices $\mathbf{X}(\boldsymbol{\lambda}, \mathbf{B})$ and  $\mathbf{Z}(\boldsymbol{\lambda}, \mathbf{B})$, and the use of automatic differentiation. Automatic differentiation has been used for fitting mixed models by several authors \citep{brooksGlmmTMBBalancesSpeed2017,fournierADModelBuilder2012,kristensenTMBAutomaticDifferentiation2016,skaugAutomaticDifferentiationFacilitate2002,skaugAutomaticApproximationMarginal2006}, but we are not aware of previous use of this technique for estimating models with factor structures.

\section{Conclusion}
\label{sec:conclusion}

We have introduced generalized additive latent and mixed models, for multilevel modeling with latent and observed variables depending smoothly on observed variables. We have also proposed an algorithm for estimating the models which scales well with large and complex data. The work was motivated by applications in cognitive neuroscience, and we have presented two examples in which the proposed models enabled new analyses not easily performed with currently available tools.

\bigskip
\begin{center}
{\large\bf SUPPLEMENTARY MATERIAL}
\end{center}

\begin{description}

\item[Online Resource 1] Additional figures and tables to the application example and simulation experiments described in Section \ref{sec:latent_response}. (pdf document)
\item[Online Resource 2] Additional figures and tables to the application example and simulation experiments described in Section \ref{sec:latent_covariates}. (pdf document)
\item[Online Resource 3] R package \verb!galamm! implementing the methods. (available from\\ https://github.com/LCBC-UiO/galamm)
\item[Online Resource 4] R scripts for analyses and simulation experiments. (available from\\ https://github.com/LCBC-UiO/galamm-scripts)
\end{description}

\bibliography{manuscript.bbl}
\end{document}


\author{}
\date{}

\maketitle

Table \ref{tab:cog_age} shows the posterior probability of each possible ordering of the age at maximum across the three cognitive domains. Tables \ref{tab:cognition_boostrap_betas} 
and \ref{tab:cognition_boostrap_lambdas} show Wald type confidence intervals and bootstrap based confidence intervals for all regression coefficients and factor loadings, respectively. Figure \ref{fig:cog_data_dist} shows the empirical distributions of baseline age, measurement intervals, and number of timepoints in the data. In addition to being analyzed in Section 5.1, these data were also used in the simulations in Section 5.2. Figure \ref{fig:cog_sim_smooth_bias} shows the average estimated trajectories across bootstrap simulations, the true trajectory, as well as all bootstrap estimates. Figure \ref{fig:edf} shows a histogram of the sum of effective degrees of freedom of all three trajectories across bootstrap samples.

\begin{table}
\centering
\begin{threeparttable}
\caption{Probability of each possible ordering of age at maximum for the three cognitive abilities studied in Section 5.1.}
\label{tab:cog_age}
\small
\begin{tabular}{ll}
\toprule
Ordering (by increasing age at maximum ability) & Probability \\
\midrule
Episodic memory $<$ Executive function $<$ Working memory & 1.62 \% \\
Episodic memory $<$ Working memory $<$ Executive function & 2.87 \% \\
Executive function $<$ Episodic memory $<$ Working memory & 88.4 \%  \\
Executive function $<$ Working memory $<$  Episodic memory & 6.82 \%  \\
Working memory $<$ Episodic memory $<$  Executive function & 0.28 \%  \\
Working memory $<$ Executive function $<$  Episodic memory & $<0.1$ \%  \\
\bottomrule
\end{tabular}
\end{threeparttable}
\end{table}

\begin{table}
\centering
\begin{threeparttable}
\caption{Regression coefficients in latent response model from Section 5. The second column contains point estimates and asymptotic standard errors and the third column contains bootstrap estimates and standard errors.}
\label{tab:cognition_boostrap_betas}
\small
\begin{tabular}{lll}
\toprule
Parameter & Estimate (95\% CI) & Bootstrap (95\% CI) \\
\midrule
\multicolumn{3}{l}{\textit{Episodic memory}}\\
\hspace{3mm} CVLT trial 1 $\beta_{t1}$ & -0.26 (-0.29, -0.23)  & -0.25 (-0.28, -0.22) \\
\hspace{3mm} CVLT trial 2 $\beta_{t2}$ & 0.74 (0.71, 0.78)  & 0.77 (0.73, 0.81) \\
\hspace{3mm} CVLT trial 3 $\beta_{t3}$ &  1.42 (1.37, 1.47)  & 1.46 (1.40, 1.50) \\
\hspace{3mm} CVLT trial 4 $\beta_{t4}$ & 1.84 (1.78, 1.89) & 1.88 (1.75, 1.88)\\
\hspace{3mm} CVLT trial 5 $\beta_{t5}$ & 2.17 (2.11, 2.24)  & 2.22 (2.16, 2.28) \\
\hspace{3mm} CVLT 5 min delay $\beta_{t6}$ & 1.62 (1.56, 1.67) & 1.66 (1.60, 1.72) \\
\hspace{3mm} CVLT 30 min delay $\beta_{t7}$ & 1.81 (1.75, 1.88) & 1.87 (1.80, 1.93) \\
\midrule
\multicolumn{3}{l}{\textit{Working memory}} \\
\hspace{3mm} Digit span backward $\beta_{t8}$ &  -0.39 (-0.41, -0.36) & -0.39 (-0.41, -0.36) \\
\hspace{3mm} Digit span forward  $\beta_{t9}$ & 0.29 (0.26, 0.31) & 0.29 (0.26, 0.31)\\
\midrule
\multicolumn{3}{l}{\textit{Executive function} (units = seconds)} \\
\hspace{3mm} Stroop 1 $\beta_{t10}$ & 32.2 (31.8, 32.5) & 32.1 (31.7, 32.4) \\
\hspace{3mm} Stroop 2  $\beta_{t11}$ & 23.3 (22.9, 23.6) & 23.2 (22.9, 23.6) \\
\hspace{3mm} Stroop 3 $\beta_{t12}$ & 58.3 (57.6, 59.0) & 58.3 (57.6, 58.9) \\
\hspace{3mm} Stroop 4 $\beta_{t13}$ & 65.4 (64.7, 66.1) & 65.3 (64.7, 66.0) \\
\midrule
\multicolumn{3}{l}{\textit{Retest effects}}  \\
\hspace{3mm} CVLT $\beta_{r1}$ & 0.11 (0.08, 0.15) & 0.12 (0.08, 0.15) \\
\hspace{3mm} Digit span $\beta_{r2}$ & 0.08 (0.05, 0.11) & 0.08 (0.05, 0.11) \\
\hspace{3mm} Stroop conditions 1 and 2  $\beta_{r3}$ & -1.24 (-0.79, -1.69) & -1.23 (-0.83, -1.65) \\
\hspace{3mm} Stroop conditions 3 and 4  $\beta_{r4}$ & -2.21 (-1.51, -2.90) & -2.24 (-1.50, -2.87) \\
\bottomrule
\end{tabular}
\end{threeparttable}
\end{table}

\begin{table}
\centering
\begin{threeparttable}
\caption{Factor loadings in latent response model from Section 5. The second column contains point estimates and asymptotic standard errors and the third column contains bootstrap estimates and standard errors. Fixed factor loadings are omitted since their uncertainty by definition is zero.}
\label{tab:cognition_boostrap_lambdas}
\small
\begin{tabular}{lll}
\toprule
Parameter & Estimate (95\% CI) & Bootstrap (95\% CI) \\
\midrule
\multicolumn{3}{l}{\textit{Episodic memory}}\\
\hspace{3mm} CVLT trial 2 $\lambda_{12}$ & 1.79 (1.72, 1.85) & 1.77 (1.68, 1.85) \\
\hspace{3mm} CVLT trial 3 $\lambda_{13}$ & 2.44 (2.35, 2.52) & 2.40 (2.28, 2.53) \\
\hspace{3mm} CVLT trial 4 $\lambda_{14}$ & 2.76 (2.66, 2.87) & 2.72 (2.59, 2.86) \\
\hspace{3mm} CVLT trial 5 $\lambda_{15}$ & 3.02 (2.90, 3.13) & 2.97 (2.83, 3.13) \\
\hspace{3mm} CVLT 5 min delay $\lambda_{16}$ & 3.04 (2.93, 3.15) & 2.99 (2.85, 3.13) \\
\hspace{3mm} CVLT 30 min delay $\lambda_{17}$ & 3.22 (3.10, 3.34) & 3.17 (3.02, 3.33) \\
\midrule
\multicolumn{3}{l}{\textit{Working memory}} \\
\hspace{3mm} Digit span forward $\lambda_{22}$ & 0.96 (0.90, 1.02) & 0.96 (0.89, 1.03) \\
\midrule
\multicolumn{3}{l}{\textit{Executive function} (units = seconds)} \\
\hspace{3mm} Stroop 2 $\lambda_{32}$ & -3.96 (3.51, 4.41) & -3.95 (3.50, 4.46) \\
\hspace{3mm} Stroop 3 $\lambda_{33}$ & -20.2 (19.3, 21.1) & -21.2 (19.8, 22.7) \\
\hspace{3mm} Stroop 4 $\lambda_{34}$ & -21.7 (20.7, 22.7) & -22.8 (21.3, 24.4) \\
\bottomrule
\end{tabular}
\end{threeparttable}
\end{table}

\begin{figure}
\centering
\includegraphics[width=\columnwidth]{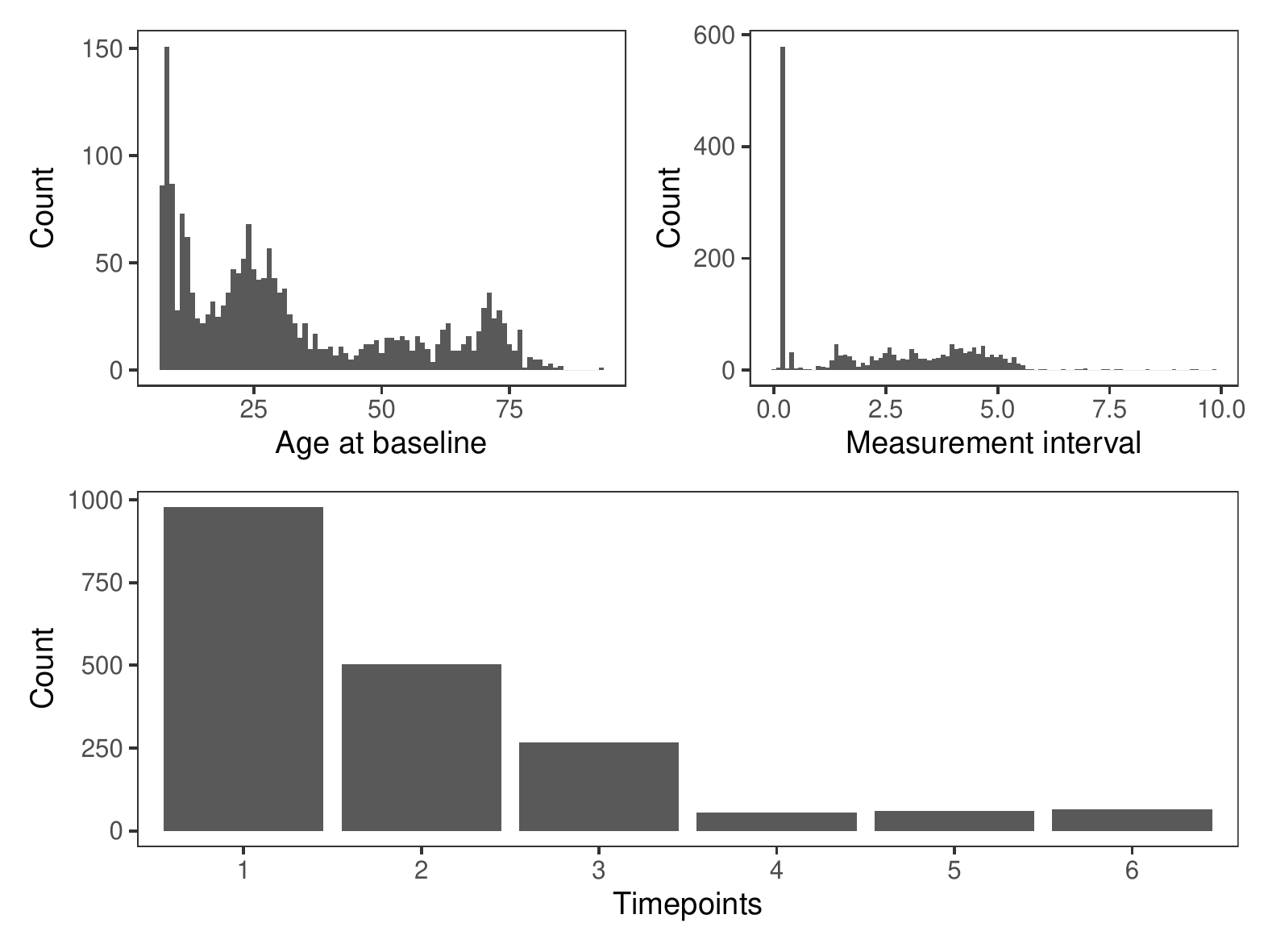}
\caption{\textbf{Age, interval, and timepoint distributions in data.} The top left plot shows a histogram of baseline age and the top right plot a histogram of times between measurements used in the data used in Section 5. The bottom plot shows the numbers of timepoints per participant.}
\label{fig:cog_data_dist}
\end{figure}

\begin{figure}
\centering
\includegraphics[width=\columnwidth]{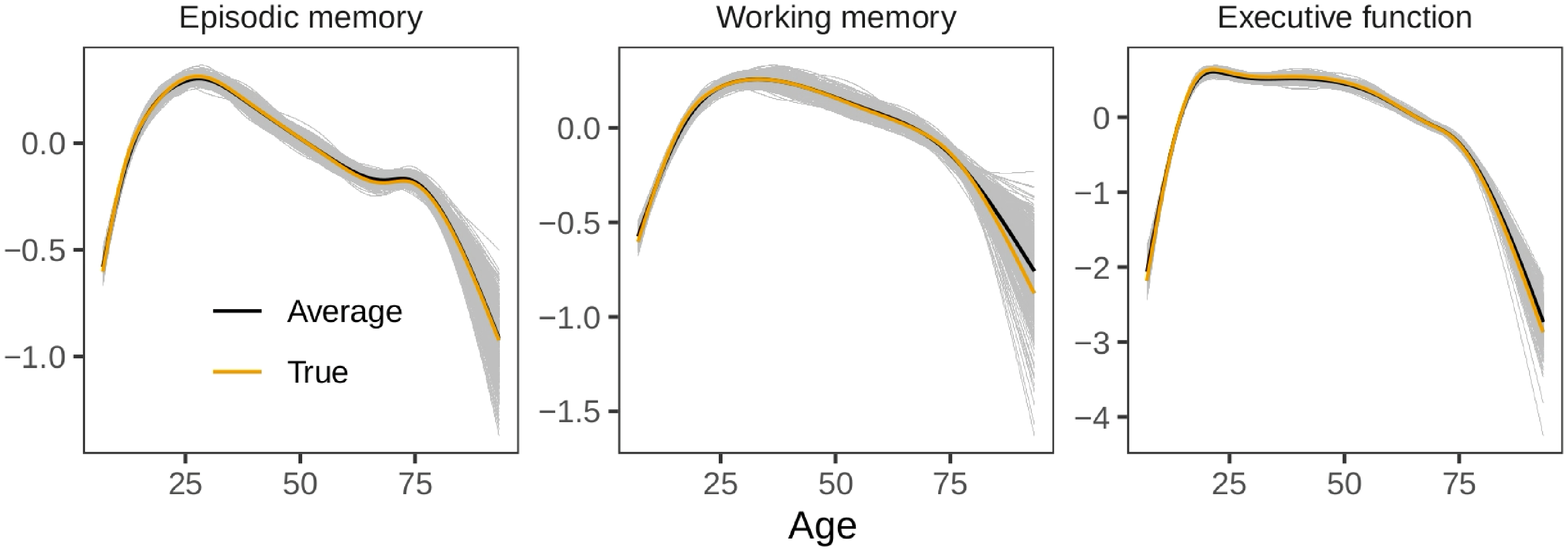}
\caption{\textbf{Results of simulation study.} Average estimates across bootstrap samples plotted together with the true curve for each cognitive domain. Gray curves in the background show all bootstrap estimates.}
\label{fig:cog_sim_smooth_bias}
\end{figure}

\begin{figure}
\centering
\includegraphics[width=\columnwidth]{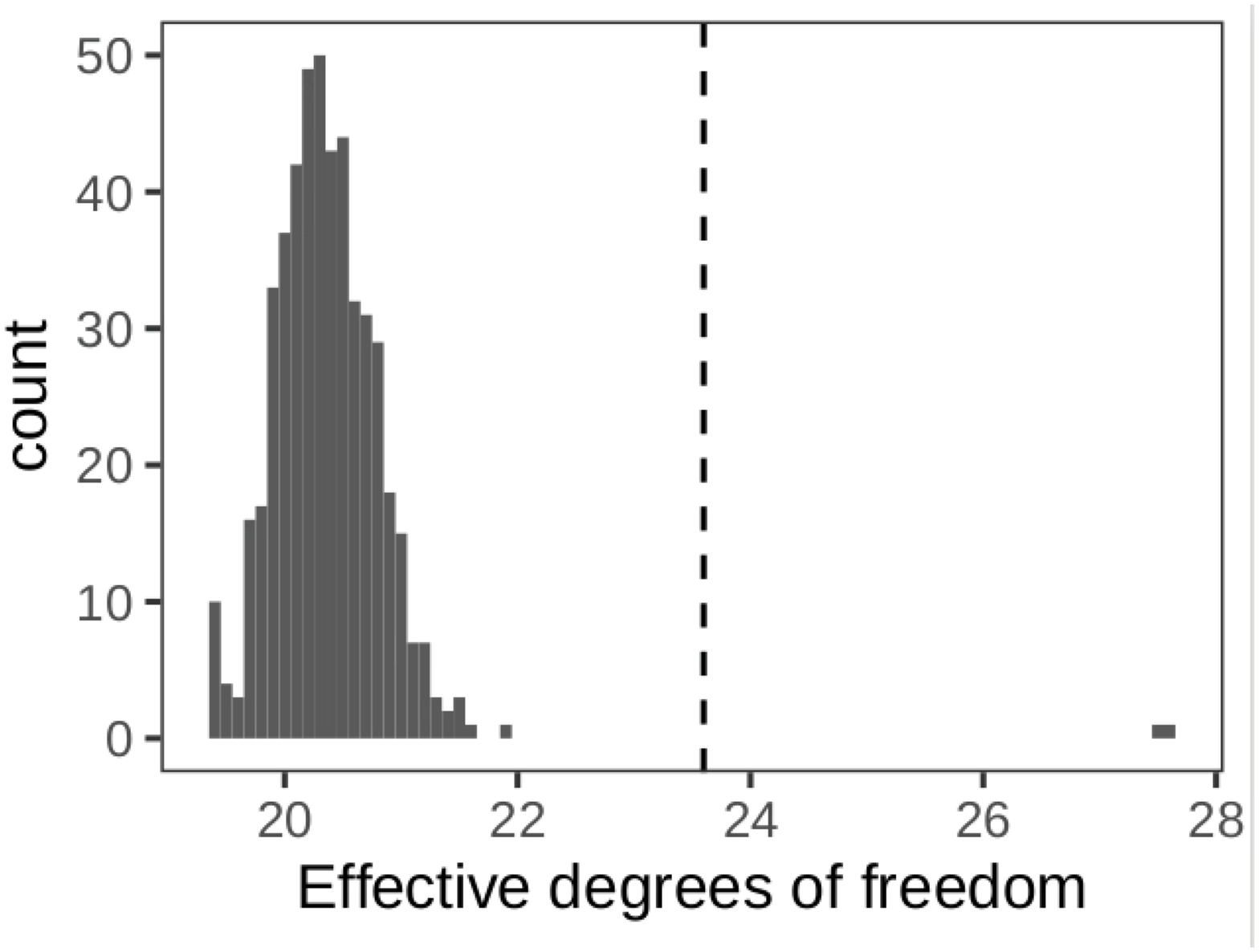}
\caption{\textbf{Effective degrees of freedom of bootstrap estimates.} Histogram of total effective degrees of freedom for the three curves across all 500 bootstrap samples. The dashed line shows the effective degrees of freedom of the model estimates, which was used as a data generating mechanism.}
\label{fig:edf}
\end{figure}


\author{}
\date{}

\maketitle

\begin{figure}
\centering
\includegraphics[width=\columnwidth]{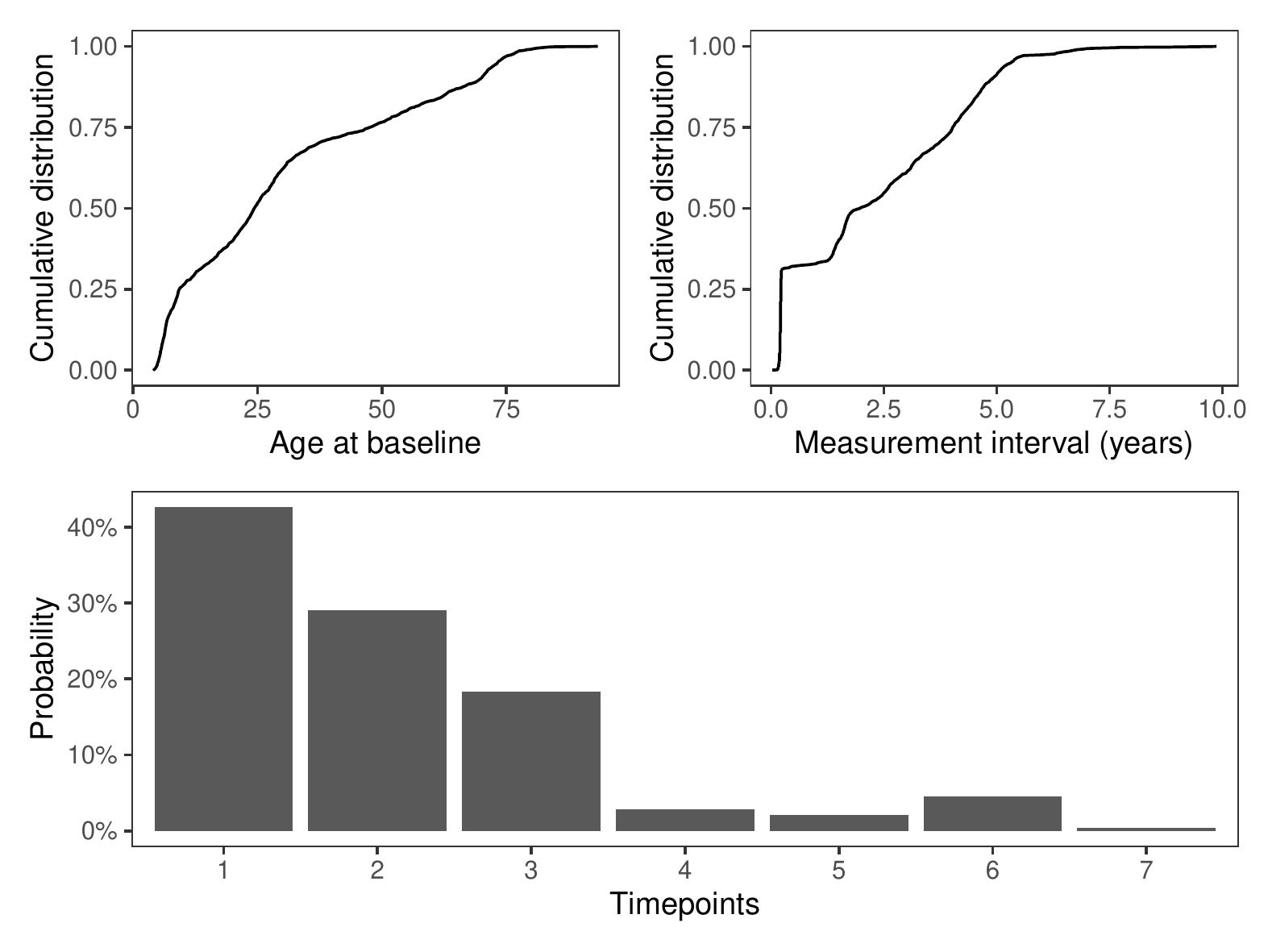}
\caption{\textbf{Age, interval, and timepoint distributions used in simulations.} The top left plot shows the cumulative distribution of baseline age and the top right plot the cumulative distribution of times between measurements used in the simulations described in Section 6.2. The bottom plot shows the probability mass function over numbers of timepoints per participant.}
\label{fig:ses_sim_dist}
\end{figure}

\begin{figure}
\centering
\includegraphics[width=\columnwidth]{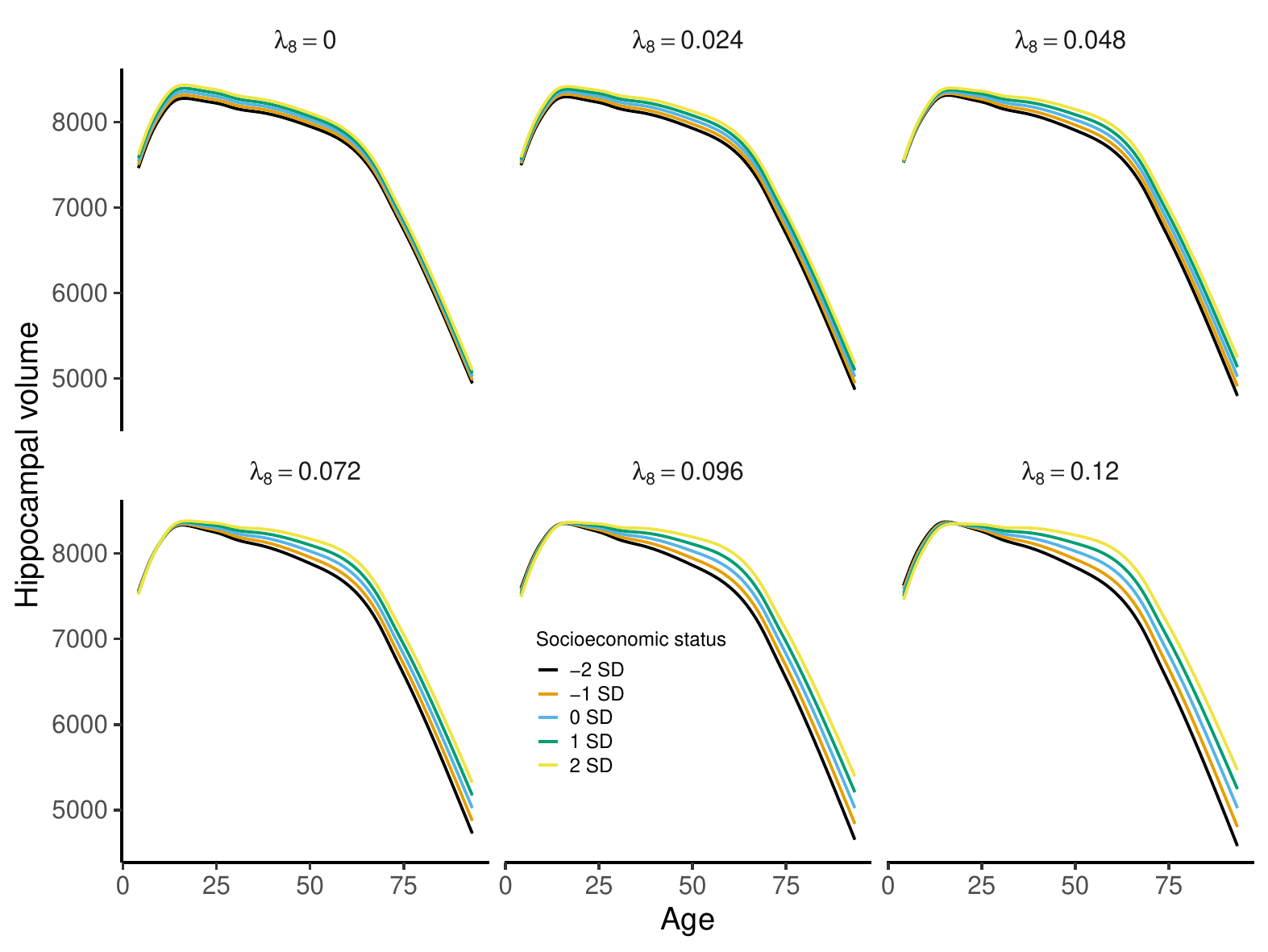}
\caption{\textbf{Interaction values used in simulation experiments.} Visualization of the six interaction values $\lambda_{8}$ used in simulation experiments described in Section 6.2.}
\label{fig:ses_sim_interaction}
\end{figure}

\begin{figure}
\centering
\includegraphics[width=\columnwidth]{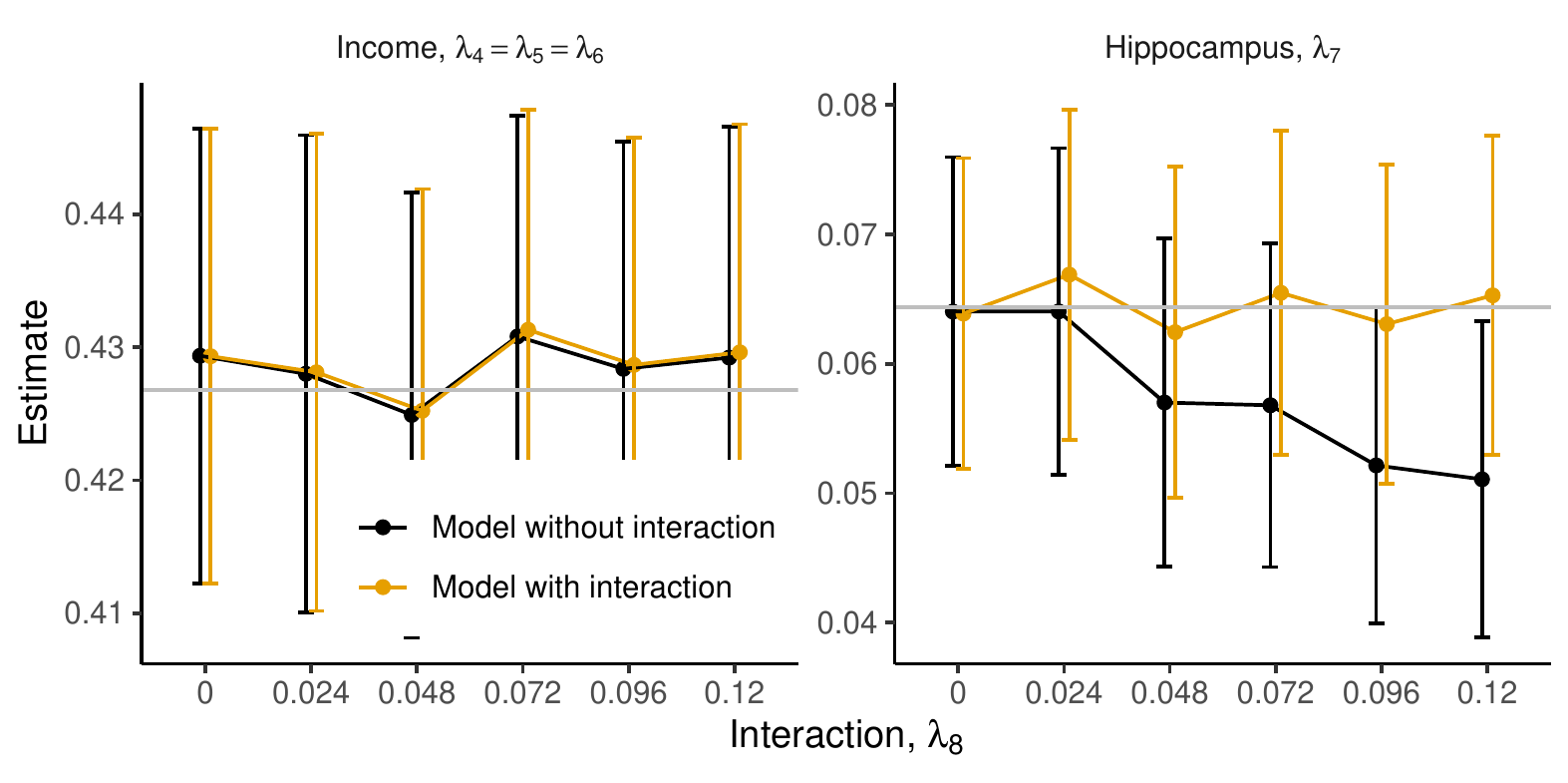}
\caption{\textbf{Estimates of factor loadings in simulation experiments.} Estimated factor loadings in simulations with latent covariates described in Section 6.2. Horizontal gray lines show the true values of the parameter, and error bars show 95\% confidence intervals due to Monte Carlo error.}
\label{fig:ses_sim_lambda_estimates}
\end{figure}

Table \ref{tab:ses_parametric_terms} shows item intercepts for the latent covariates model presented in Section 6.1. Figure \ref{fig:ses_sim_dist} shows the distributions of baseline age, measurement intervals, and number of timepoints per participant, used in the simulations in Section 6.2. Figure \ref{fig:ses_sim_interaction} visualizes the trajectories under the interaction effects assumed in the simulation studies. Figure \ref{fig:ses_sim_lambda_estimates} shows bias of estimated factor loadings.

\begin{table}
\centering
\begin{threeparttable}
\caption{Item intercepts in model of hippocampal volume and socioeconomic status.}
\label{tab:ses_parametric_terms}
\begin{tabular}{lllll}
\toprule
Parameter & Estimate & SE & Units \\
\midrule
\multicolumn{4}{l}{\textit{Item intercepts}}\\
\hspace{3mm} Father's education, $\beta_{s1}$ & 2.81 & 0.00684 & $\log(\text{years})$ \\
\hspace{3mm} Mother's education, $\beta_{s2}$ & 2.81 & 0.00671 & $\log(\text{years})$ \\
\hspace{3mm} Education, $\beta_{s3}$ & 2.81 & 0.0053 & $\log(\text{years})$ \\
\hspace{3mm} Father's income, $\beta_{s4}$ & 13.2 & 0.0399 & $\log(\text{NOK})$ \\
\hspace{3mm} Mother's income, $\beta_{s5}$ & 13.2 & 0.0368 & $\log(\text{NOK})$ \\
\hspace{3mm} Income, $\beta_{s6}$ & 13.1 & 0.0367 & $\log(\text{NOK})$ \\
\bottomrule
\end{tabular}
\end{threeparttable}
\end{table}

